\documentclass[letterpaper,english,reprint, aps]{revtex4-1}
\usepackage[LGR,T1]{fontenc}
\usepackage[latin9]{inputenc}
\usepackage{array}
\usepackage{textcomp}
\usepackage{multirow}
\usepackage{amsmath}
\usepackage{amssymb}
\usepackage{graphicx}

\makeatletter

\pdfpageheight\paperheight
\pdfpagewidth\paperwidth

\DeclareRobustCommand{\greektext}{%
  \fontencoding{LGR}\selectfont\def\encodingdefault{LGR}}
\DeclareRobustCommand{\textgreek}[1]{\leavevmode{\greektext #1}}
\ProvideTextCommand{\~}{LGR}[1]{\char126#1}

\providecommand{\tabularnewline}{\\}

\makeatother

\usepackage{babel}
\begin{document}
\title{Studying Same-Sign Top Pair Production Through Top-Higgs FCNC Interactions
at the HL-LHC}
\author{O. M. Ozsimsek}
\email{ozgunozsimsek@hacettepe.edu.tr}

\affiliation{Hacettepe University, Institute of Science, Beytepe, 06800 Ankara,
Turkey}
\author{V. Ari}
\email{vari@science.ankara.edu.tr}

\affiliation{Department of Physics, Ankara University, 06100 Ankara, Turkey}
\author{O. Cakir}
\email{ocakir@science.ankara.edu.tr}

\affiliation{Department of Physics, Ankara University, 06100 Ankara, Turkey}
\date{\today}
\begin{abstract}
We investigate the potential of the HL-LHC for discovering new physics
effects via the same-sign top pair signatures. We focus on the semi
leptonic (electron and muon) decay of the top quarks and study the
reach for a simplified model approach where top quark flavor changing
could occur through a neutral scalar exchange. A relatively smaller
background contribution and clean signature are the advantages of
the leptonic decay mode of the same-sign $W$ bosons in the same-sign
production processes of top quark pairs. Assuming the FCNC between
top quark, up type quark and scalar boson from the new physics interactions
the branchings could be excluded of the order ${\cal O}(10^{-4})$.
We use angular observables of the same-sign lepton pairs and the top
quark kinematics in the process which provide the possibility of separation
of new physics signal from the SM backgrounds using machine learning
teqniques. We find that the same-sign top quark pair production is
quite capable of testing the top-Higgs FCNCs at the HL-LHC. 

\emph{Our work is presented in Arxiv with preprint number: arXiv:2105.03982.}
\end{abstract}
\pacs{1234}
\maketitle

\section{Introduction}

Among all fundamental fermions in the standard model (SM), top quark
has the largest mass and causes the most serious hierarchy, and plays
an essential role in the metastability of the Higgs boson potential
\citep{key-1}. Top quark is also the last corner stone of the family
structure of the SM with a huge mass gap with other members of quark
content of the SM. It is the most sensitive particle for TeV scale
physics in SM with Higgs boson, therefore researching the interactions
of top quark is a crucial part of BSM physics.

The flavor changing neutral currents (FCNCs) among the up or down
sector quarks are not present at leading-order in both Yukawa and
gauge interactions within the standard model (SM) framework. However,
extremely small FCNC couplings could be generated from loop-level
diagrams which are strongly suppressed due to the Glashow-Iliopoulos-Maiani
(GIM) mechanism \citep{key-2} and it is one of the unique characteristics
of the SM. Besides it sets a new horizon for new researches.

The essence and importance of GIM mechanism's veto and studying FCNC
interactions lies in the decision of dropping or keeping the FCNC
preventing unique feature of SM model to new physics. FCNC searches
will deduce its ultimate fate without any doubt. If one can show its
possibility, that would be a great progress at BSM researches.

The phenomenology of FCNC couplings has been discussed in many studies.
There are scenarios including top-Higgs FCNC within supersymmetry
models (including MSSM and RPV) \citep{key-32,key-34,key-3,key-46},
the two Higgs doublet models (both flavor violating and conserving)
\citep{key-31,key-33,key-4,key-51}, quark singlet models \citep{key-30},
composite Higgs models \citep{key-35} and warped extra dimensions
models \citep{key-36}. However, we use an effective lagrangian formalism
for top-Higgs-q ($thq$) FCNC interactions \citep{key-14,key-15}
for a model independent research and discuss our results with the
expectations of present models. Then, top quark FCNCs can also show
up in the processes through the exchange of a new neutral scalar.
To be specific about the branching ratios models are expected we would
like to summarize them at a Table \ref{tab:Expected-branching-ratios}.
\begin{table}[h]
\caption{Expected FCNC branching ratios from models \label{tab:Expected-branching-ratios}\citep{key-42}.}

\begin{tabular}{|c|c|c|}
\hline 
 & $t\to uH$ & $t\to cH$\tabularnewline
\hline 
SM & $2\times10^{-17}$ & $3\times10^{-15}$\tabularnewline
\hline 
QS & $4.1\times10^{-5}$ & $4.1\times10^{-5}$\tabularnewline
\hline 
MSSM & $\leq10^{-5}$ & $\leq10^{-5}$\tabularnewline
\hline 
2HDM (FC) & $-$ & $\leq10^{-5}$\tabularnewline
\hline 
2HDM(FV) & $6\times10^{-6}$ & $2\times10^{-3}$\tabularnewline
\hline 
RPV SUSY & $\leq10^{-9}$ & $\leq10^{-9}$\tabularnewline
\hline 
\end{tabular}
\end{table}

Production of two positively charged top quarks via $uu\rightarrow tt$
resulting in an excess of same-sign lepton pairs have already been
searched by the ATLAS \citep{key-49}. Systematic uncertainties for
the main backgrounds, including charge misidentification, fake/non-prompt
leptons, etc. are presented about 28\% , 33\% and 30\% for the $ee$
,$e\mu$ ,$\mu\mu$ channels, respectively. The limit on the cross
section leads to a limit of $BR(t\rightarrow uH)<0.01$. Another search
for flavour changing neutral current processes in top quark decays
have been presented again by the ATLAS Collaboration from proton-proton
collisions at the LHC with $\sqrt{s}=13$ TeV \citep{key-5}. The
observed (expected) upper limits are set on the $t\to cH$ branching
ratio of $1.1\times10^{-3}$ ($8.3\times10^{-4}$) and on the $t\to uH$
branching ratio of $1.2\times10^{-3}$ ($8.3\times10^{-4}$) at the
$95\%$ confidence level. A search for flavor-changing neutral currents
(FCNC) in events with the top quark and the Higgs boson is presented
by the CMS collaboration \citep{key-6}. The observed (expected) upper
limits at $95\%$ confidence level are set on the branching ratios
of top quark FCNC decays, $BR(t\to uH)<7.9\times10^{-4}$ ($1.1\times10^{-3}$)
and $BR(t\to cH)<9.4\times10^{-4}$ ($8.6\times10^{-4}$), assuming
a single non-zero FCNC coupling. These prior works are the basically
constitutes the starting point of FCNC researches at post Higgs era
researches for FCNC. More recently, the limits given by CMS collaboration
have been further improved as $BR(t\to uH)<1.9\times10^{-4}$ $(3.1\times10^{-4})$
and $BR(t\to cH)<7.3\times10^{-4}$ $(5.1\times10^{-4})$ \citep{key-92}. 

Especially we focus on the limits by CMS collaboration \citep{key-92}:
we use same effective Lagrangian up to a factor of weak coupling constant
$g$ and follow their limits on coupling constants which are 0.037
for $\eta_{u}$ and 0.071 for $\eta_{c}$. In the case of ATLAS collaboration
\citep{key-5}, they give an lower bound arround 0.065 for both coupling
constants. However, bear in mind that the differences mainly come
from using different effective Lagrangian which is discussed at model
framework section. We investigate the problem by setting coupling
constant 0.07 for scenarios about to introduce at following sections
and try to improve the limits.

In recent years many new collider ideas such as HL-LHC/HE-LHC/FCC
\citep{key-7,key-8,key-9} have been reported and technical design
report of HL-LHC has been published. Most promising feature of the
HL-LHC collider for BSM searches is increased COM energy ($14$ TeV)
and especially its luminosity \citep{key-7} of up to $3\ ab^{-1}$.
Some phenomenological researches for future colliders and HL-LHC have
already been started up; for $tqH$ couplings have been explored at
a high-luminosity $ab^{-1}$ $ep$ colliders (with the possibility
of electron beam having a polarization of 80\% and electron energy
is typical 60 GeV), the $2\sigma$ upper limits on $Br(t\rightarrow uH)$
have been obtained as $1.5\times10^{-3}$ and $2.9\times10^{-4}$
at the future colliders LHeC and FCC-eh, respectively \citep{key-91}.

Development of such a collider has notable effects on the BSM literature
evidently since it to offer new possibilities for phenomenological
studies and gives a large room for potential discoveries/exclucions.
It offers an opportunity to rule out flavor violating 2HDM for $t\to cH$
case and penetrate the other regions forseen by other models (such
as Randall-Sundrum model) \citep{key-37}. As a consequence exploiting
the physics potential of HL-LHC is crucial for next phase of BSM searches.

\begin{figure}[h]
\begin{raggedright}
\includegraphics[scale=0.48]{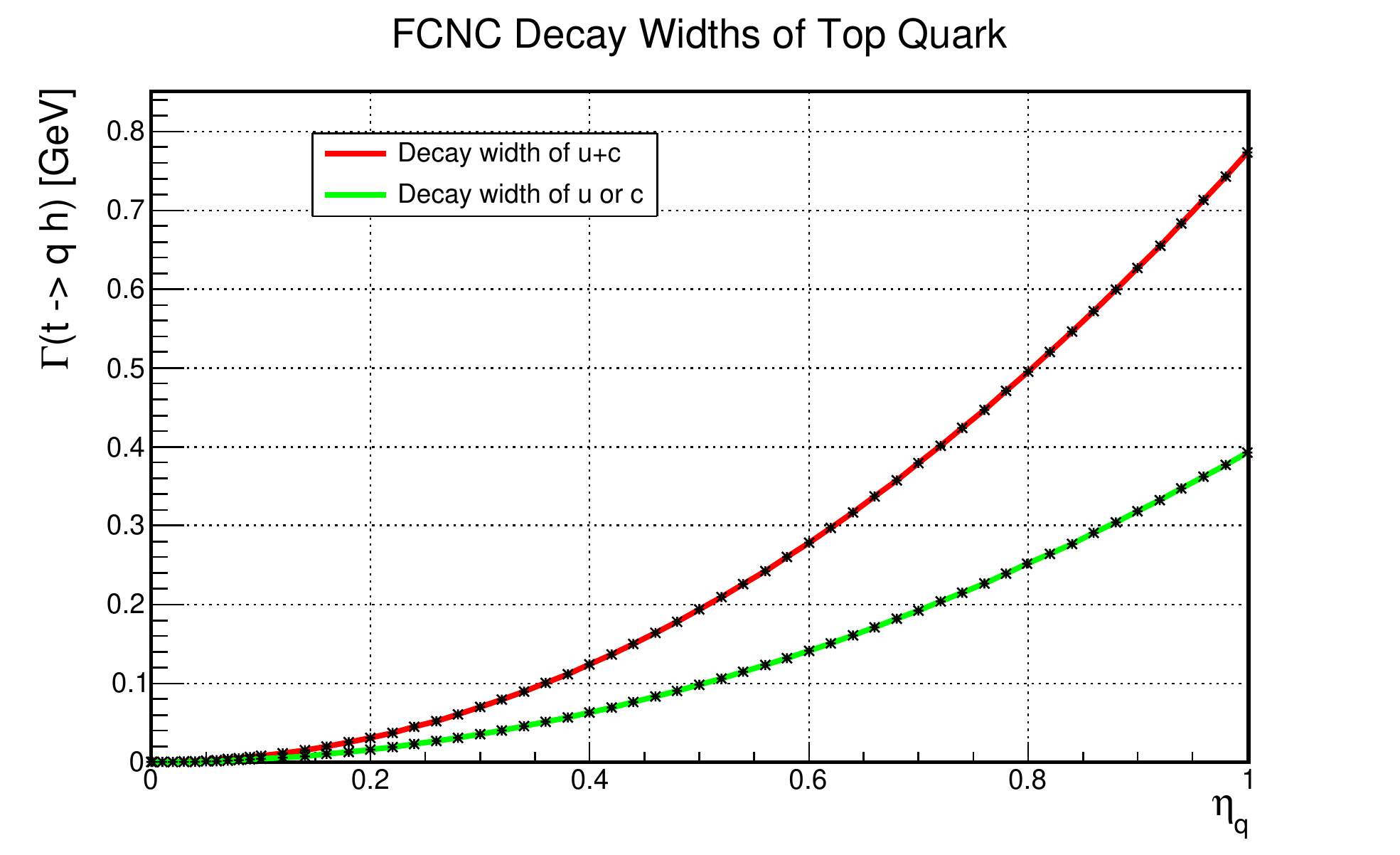}
\par\end{raggedright}
\raggedright{}\caption{The FCNC decay width of top quark according to two scenarios: For
$u+c$ case top quark can decay into both, otherwise decays into only
one of them.\label{fig:The-FCNC-decay}}
\end{figure}

\begin{figure}[h]
\begin{raggedright}
\includegraphics[scale=0.48]{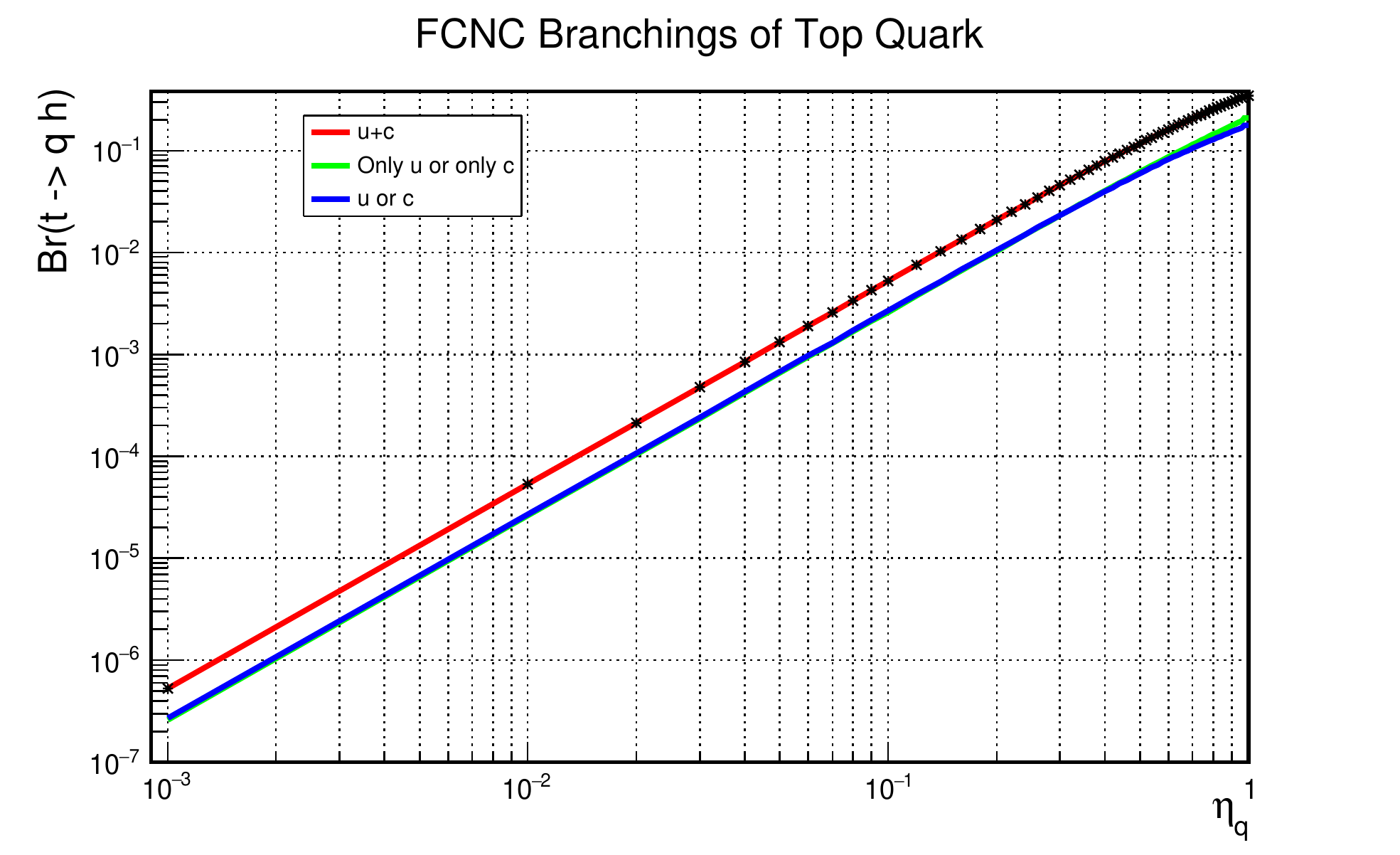}
\par\end{raggedright}
\raggedright{}\caption{FCNC branchings of top quark for three scenario presented: $u+c$
indicates top quark can decay them both via Higgs. Only $u$ or only
$c$ assumes top quark can decay only one of them and the other channel
is closed. At last case top both channels are accesible but only one
of them preferred. To access beyond the FCNC regions excluded by LHC
$(\mathrm{Br}(t\rightarrow qh)\sim\mathcal{O}(10^{-3}))$, one needs
to set bound for the coupling constant roughly below $\eta_{q}=0.04$.\label{fig:FCNC-branchings-of}}
\end{figure}

The phenomenological researches and simulations based on new colliders
started to making predictions about new physics scenarios and set
new limitations. To be specific at HL-LHC for FCNC interactions \citep{key-10,key-11,key-12,key-13,key-40},
branching ratios are updated as $BR(t\rightarrow qh)<\mathcal{O}(10^{-4})$
using various different analyses from different channels and processes;
thus couplings are expected to go below $\eta_{q}=0.04$ which is
rougly below the known limits from experiments. Expected FCNC decay
widths and branching ratios are given at Fig. \ref{fig:The-FCNC-decay}
and \ref{fig:FCNC-branchings-of} respectively according to three
scenarios which are important and handled seperately to set limits
for couplings in following sections.

In this study, we would like to investigate the problem and seek for
the new limits at HL-LHC. To do so, we restrict ourselves to production
mechanisms of same sign $tt(\bar{t}\bar{t})$ pairs (signal processes
$pp\to tt\to W^{+}W^{+}bb\to l^{+}l^{+}bb+MET$ , $pp\to\bar{t}\bar{t}\to W^{-}W^{-}\bar{b}\bar{b}\to l^{-}l^{-}\bar{b}\bar{b}+MET$)
including the exchange of Higgs boson at the HL-LHC. In addition,
because the analysis was carried out in the HL-LHC, with the anticipation
that the systematic uncertainties described in the literature would,
in general, reduce, a value of 20\% was used for the systematic uncertainties,
which is still near to the limitations given in the literature and
discussed in the findings section. This was done in light of the fact
that the analysis was carried out in the HL-LHC.We introduce the kinematical
variables to enhance the signal (S) and background (B) ratio. Angular
separation of the two same sign leptons could indicate the new physics
effects in $tt(\bar{t}\bar{t})$ production process, and separate
the signal from background processes.

In order to make this research more detailed and similar to other
studies in the literature (for comparison purposes), three different
scenarios were designed for the signaling process. These are the u+c,
only u and only c scenarios. As the nomenclature suggests, FCNC transitions
are made possible from the top quark to the other two quarks in the
u+c case, while in other cases the transitions are limited to just
one quark. 
\begin{figure}[h]
\includegraphics[scale=0.028]{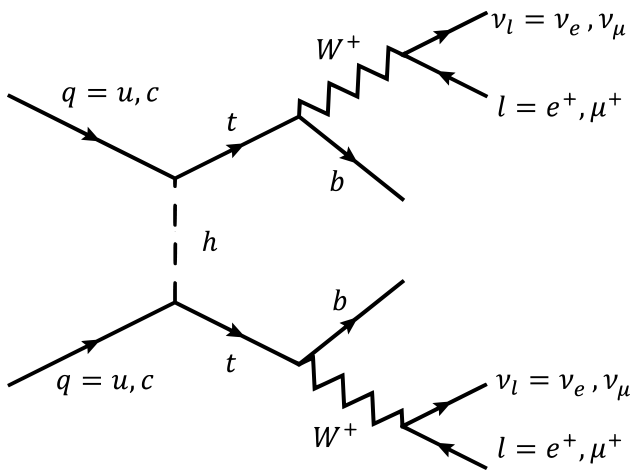}
\raggedright{}\caption{Feynman diagram of leptonic decay of top-Higgs FCNC process: For sake
of simplicity we give a compact form of the process which includes
two new physics vertex. Dominant contribution will come from $pp\rightarrow tt$
process due to the large values of PDF of up type quarks in proton.\label{fig:Feynman-diagram-of}}
\end{figure}

\section{Model Framework}

The flavour changing neutral current interactions of the top quark
with other particles of the SM have been described in a general way
as an extension \citep{key-14,key-15}. This provides a direct connection
between experimental observables and the new anomalous couplings.
The Lagrangian describing FCNC $tqH$ interactions in model independent
manner is given as

\begin{align}
L_{H} & =\frac{1}{\sqrt{2}}H\bar{t}(\eta_{u}^{L}P^{L}+\eta_{u}^{R}P^{R})u+h.c.\nonumber \\
 & +\frac{1}{\sqrt{2}}H\bar{t}(\eta_{c}^{L}P^{L}+\eta_{c}^{R}P^{R})c+h.c.\label{eq:1}
\end{align}
where the $\eta_{q}^{L/R}$ couplings set the strength of the coupling
between the top quark, the Higgs boson and up or charm quark, as well
as the chirality of this coupling. They can be complex in general,
however we take into account real parts of the couplings to reduce
the free parameters. In literature this interaction can be seen as
modeled without the constant $\frac{1}{\sqrt{2}}$ thus gives higher
top branchings a factor of 2 \citep{key-199}. The FCNC processes
that corresponds to $tqh$ interactions have been described by a similar
Lagrangian \citep{key-6} with an extra factor of weak coupling constant.
To switch between models we just need to remind this conversion factor.
We keep that constant here in order to make bounds more strict mean
while keeping the conversion to other models in our mind. Note that
it effects cross section and number of evens naturally too, thus makes
signal processes  even harder and realistic. The decay width for FCNC
channels can be calculated as 

\begin{equation}
\Gamma(t\to qh)=\frac{(\eta_{qL}^{2}+\eta_{qR}^{2})}{64\pi}\frac{(m_{t}^{2}-m_{h}^{2})^{2}}{m_{t}^{3}}
\end{equation}
and its numerical value depends on the coupling values related to
$\Gamma(t\to qh)\simeq0.1904(\eta_{qL}^{2}+\eta_{qR}^{2})$ GeV. The
branching ratio to an FCNC channel can be expressed as $BR(t\to qh)=\Gamma(t\to qh)/\Gamma(t\to all)$.
Since the dominant decay mode of top quark is $\Gamma(t\to Wb)$,
this branching ratio mostly related to ($\eta_{qL}^{2}+\eta_{qR}^{2}$)
factor especially for smaller coupling values.

The model framework can also be compared with the formalism assumed
that the FCNC interactions occur via a weak sector. The relevant effective
interaction Lagrangian including a new flavor changing scalar ($\phi$)
is given

\begin{equation}
L_{\phi}=\phi\bar{t}(a_{u}+b_{u}\gamma^{5})u+\phi\bar{t}(a_{c}+b_{c}\gamma^{5})c+H.c.\label{eq:2}
\end{equation}
where the coupling parameters $a_{u,c}$ and $b_{u,c}$ denote the
scalar and axial couplings between top quark and up-type light quarks
($u,c$) which proceeds through the exchange of a scalar $\phi$.
To compare different formalism for the top-scalar FCNC we find the
correspondance of the couplings $a_{q}=(\eta_{q}^{L}+\eta_{q}^{R})/2\sqrt{2}$
and $b_{q}=(\eta_{q}^{R}-\eta_{q}^{L})/2\sqrt{2}$. Assuming no specific
chirality dependence (same value for left and right handed couplings)
of the process we may set $a_{q}=\eta_{q}/\sqrt{2}$ and $b_{q}=0$.

In this study, we use a template model. The paremeters that appear
in the topFCNC\_UFO \citep{key-16,key-17} model are complex numbers
in general and their real and imaginary parts can be set manually.
In this work, we restrict ourselves to real parameters in order to
reduce the free parameters.

\section{Cross Sections of Signal and Background}

\begin{figure}[h]
\noindent \raggedright{}\includegraphics[scale=0.48]{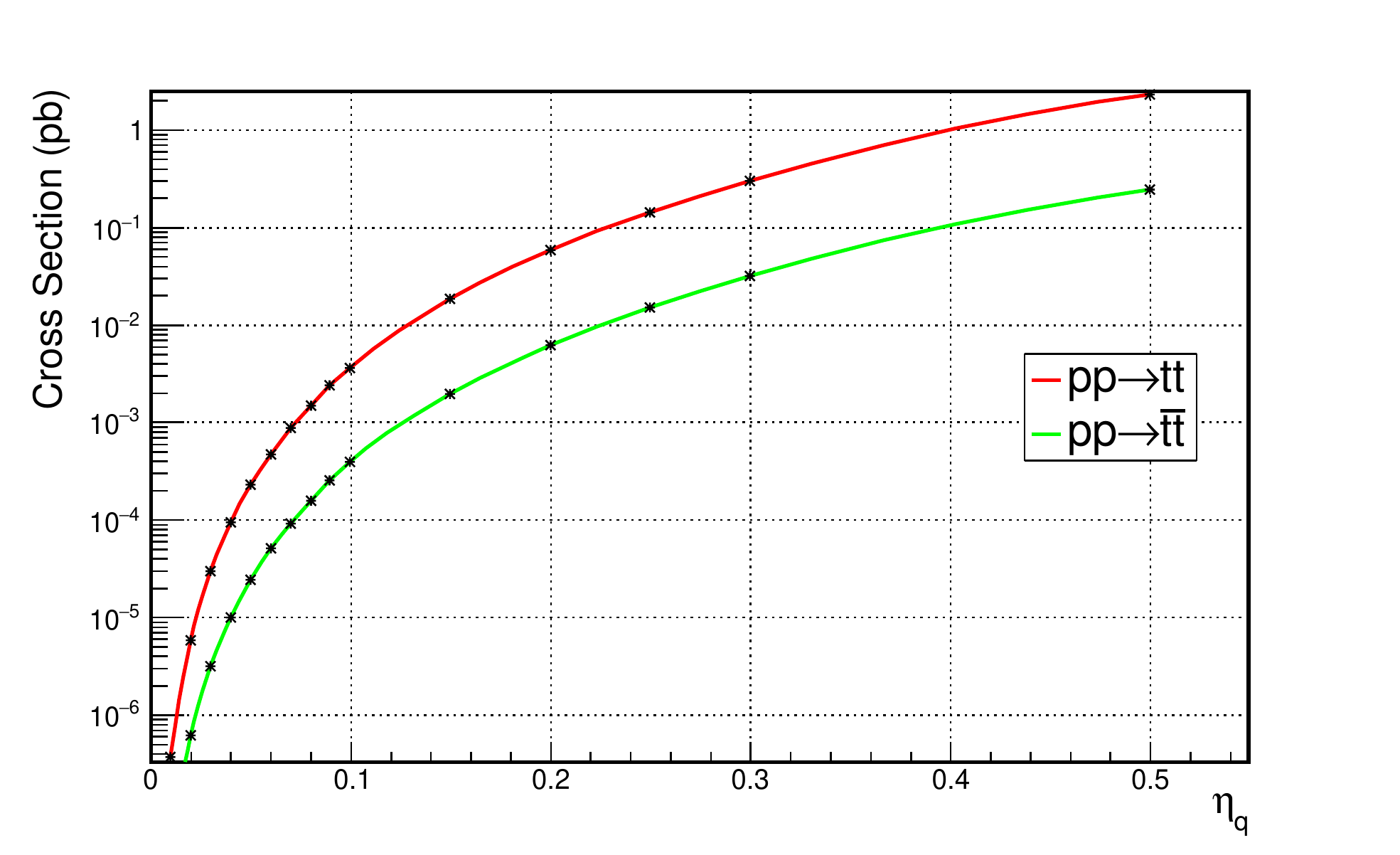}\caption{Estimated cross sections according to coupling constant of two same
sign lepton signal in FCNC processes. As we can see main contribution
comes from positively charged top pair due to higher parton distributions
of valance quarks at proton which differs nearly one order of magnitude.
Nevertheless adding $\bar{t}\bar{t}$ production we use negatively
charged lepton pair to enhance the signal process. We assume all FCNC
coefficients are the same and all channels are open (to state exactly
we use $u+c\rightarrow\eta_{u}=\eta_{c}$ case).}
\label{Figure 1}
\end{figure}

\begin{figure}[h]
\includegraphics[scale=0.48]{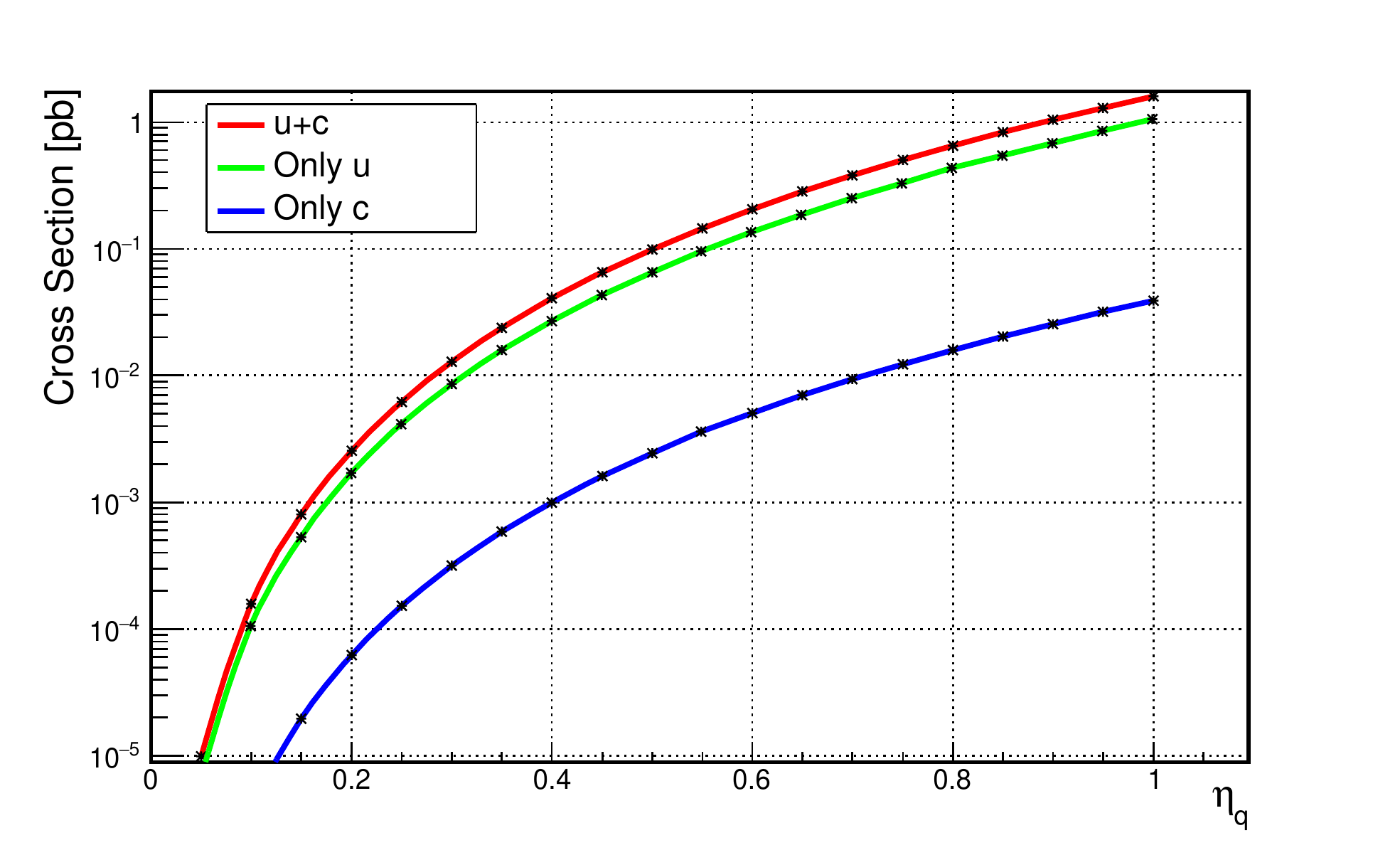}
\raggedright{}\caption{Cross section for the final state including two same sign lepton at
HL-LHC. Comparison of three FCNC scenarios: The matrix element of
the FCNC process includes two new physics vertices which are proportional
to $\eta_{q}^{2}$, hence the cross section is proportional to $\eta_{q}^{4}$.
If $c$ quark does not involve in interactions then cross section
depends only on $\eta_{u}^{4}$. Since the PDF of valence $u$ quarks
are high, the contribution to cross section from $u$ quarks are consideribly
high. In the case of a forbidden $u$ quark interaction, cross section
completely depends on $\eta_{c}^{4}$ while cross section is lower.
At $u+c$ scenario when both $u$ and $c$ actively take part at interaction
in addition to former coefficients plus additional cross terms present
and total cross section is relatively higher from only $u$ case showing
that dominant part of interactions carries the fingerprint of $u$
quark distributions in proton. \label{fig:Comparison-of-three}}
\end{figure}

At the first step before event generation we calculate the cross section
for FCNC processes including $tqh$ which vertices leads to same sign
signal final state as shown schematicaly at Fig. \ref{fig:Feynman-diagram-of}.
Since the cross-section is proportional to the modulo quartic of the
value of the anomalous couplings. In figure \ref{Figure 1} and \ref{fig:Comparison-of-three}
we can see due to presence of up type quarks in proton, $pp\rightarrow tt$
process is much more favorable than $pp\rightarrow\bar{t}\bar{t}$.
Although the contribution from the signal $pp\rightarrow\bar{t}\bar{t}$
to same sign lepton signal compared to the signal from $pp\rightarrow tt$
is nearly less than one order of magnitude, we also use that contribution
to enhance the signal.

After setting model parameters the signal samples and background samples
are generated with MadGraph5 \citep{key-18}. In the partonic and
hadronic level simulations we use the parton distribution function
(PDF) set NNPDF2.3 \citep{key-211} at MadGraph5's default energy
scale. PYTHIA8 \citep{key-19} is used for shower and hadronisation
processes and finally DELPHES 3 \citep{key-20} is used for detector
level simulation. Result files are analyzed with Root6 \citep{key-21}.

As mentioned before same sign lepton signal has relatively low backgroud,
which is advantageous and many of the background processes fall into
reducible background category which means although they are present
due to similarities between signal process by applying proper analyze
cuts their contributions can be well reduced. However there still
exist tough irreducible backgrounds. The contributions from various
backgrounds are listed below.
\begin{table}[h]
\raggedright{}\caption{Signal and background processes with leptonic decay channels\label{tab:Signal-and-background}:
We consider the positively and negatively charged leptonic final states
for maximal imitation of signal process. So background events generated
with this regard. We force particles to give $l^{+}/l^{-}$ final
states if possible. Otherwise, we let particles to decay any channel.
For $W$ boson at intermediate states we always take leptonic decay
modes for get the maximal similarity with signal.}
\begin{tabular}{|c|c|c|}
\hline 
Process & Cross section(pb) & Intermediate states\tabularnewline
\hline 
\hline 
$pp\rightarrow tt(\bar{t}\bar{t})$ &  & \multirow{4}{*}{$WWbb$}\tabularnewline
$\eta_{u}=\eta_{c}=0.07$ & $3.439\times10^{-5}$ & \tabularnewline
$\eta_{u}=0.07$ & $5.257\mathrm{\times}10^{-4}$ & \tabularnewline
$\eta_{c}=0.07$ & $1.976\times10^{-5}$ & \tabularnewline
\hline 
$pp\rightarrow t\bar{t}W^{\pm}$ & $1.647\times10^{-2}$ & $WWWbb$\tabularnewline
\hline 
$pp\rightarrow W^{\pm}W^{\pm}jj$ & $1.357\times10^{-2}$ & $WWjj$\tabularnewline
\hline 
$pp\rightarrow W^{+}W^{-}Z$ & $1.581\times10^{-3}$ & $WWZ$\tabularnewline
\hline 
$pp\rightarrow t\bar{t}l^{+}l^{-}$ & $1.827\times10^{-2}$ & $WWbbll$\tabularnewline
\hline 
$pp\rightarrow ZZW^{\pm}$ & $1.938\times10^{-4}$ & $WZZ$\tabularnewline
\hline 
$pp\rightarrow t\bar{t}W^{+}W^{-}$ & $8.466\times10^{-2}$ & $WWWWbb$\tabularnewline
\hline 
$pp\rightarrow t\bar{t}Z$ & $1.846\times10^{-4}$ & $WWbbZ$\tabularnewline
\hline 
$pp\rightarrow ZZjj$ & $1.267\times10^{-2}$ & $ZZjj$\tabularnewline
\hline 
$pp\rightarrow t\bar{t}H$ & $4.242\times10^{-5}$ & $WWbbH$\tabularnewline
\hline 
\end{tabular}
\end{table}
 
\begin{table}[h]
\raggedright{}\caption{Content of background groups: Here we grouped backgrounds to increase
clarity of our histograms. The most important feature of these backgrounds
are four of them includes top pair as backbone, and others only bosons.
Only $pp\rightarrow t\bar{t}l^{+}l^{-}$ process is left as own. That
behaviour of backgrounds lead us the catagorization of them in this
table.}
\begin{tabular}{|c|c|c|}
\hline 
Group Name & Processes & Definition\tabularnewline
\hline 
\hline 
$t\bar{t}$ w/wo boson(s) & $pp\rightarrow t\bar{t}W^{\pm}$ & Top pair with/without\tabularnewline
 & $pp\rightarrow W^{+}W^{-}t\bar{t}$ & a boson or bosons\tabularnewline
 & $pp\rightarrow t\bar{t}Z$ & \tabularnewline
 & $pp\rightarrow t\bar{t}H$ & \tabularnewline
\hline 
Bosons w/wo jets & $pp\rightarrow W^{+}W^{+}jj$ & \tabularnewline
\cline{2-2} 
 & $pp\rightarrow W^{+}W^{-}Z$ & Bosons\tabularnewline
\cline{2-2} 
 & $pp\rightarrow ZZW^{\pm}$ & with/without jets\tabularnewline
\cline{2-2} 
 & $pp\rightarrow ZZjj$ & \tabularnewline
\hline 
\end{tabular}
\end{table}

Characteristics of signal events are two jets (b-tagged if possible),
two same sign leptons, and missing transverse energy. We choose our
background processes by considering three fundamental features: 
\begin{itemize}
\item Similarity of final state particles as much as possible with signal
processes.
\item High cross section compared to signal.
\item Having same reconstruction inputs as for the signal.
\end{itemize}
Backgrounds given in Table \ref{tab:Signal-and-background} have at
least one of these properties, besides some of them have two. As long
as all processes have their own unique nature, they more or less differ
at least one criterion or partly one or two criterion.

The processes $pp\rightarrow W^{\pm}W^{\pm}jj$; $pp\rightarrow t\bar{t}W^{\pm}$;
$pp\rightarrow t\bar{t}l^{+}l^{-}$; $pp\rightarrow t\bar{t}W^{+}W^{-}$
with same-sign dilepton decay modes which are most similar to our
signal process are directly background to our signal process and they
are all irreducible. Although they give same final state content with
the signal, $pp\rightarrow t\bar{t}W^{\pm}$ reconstruction region
is slightly different. This process also gives similar products at
final state. However its cross section is high. In the case $pp\rightarrow W^{\pm}W^{\pm}jj$,
on the one hand reconstruction region is significantly different,
on the other hand its particle content is exactly the same. In addition
to previous two discussions as an advantage for analysis $pp\rightarrow t\bar{t}l^{+}l^{-}$
procces has low cross section compared to other two. Nevertheless
its reconstruction region is fairly same. $pp\rightarrow t\bar{t}W^{+}W^{-}$
with leptonic decay modes directly produce signal content, however
its reconstruction region noticeably distinct. Besides its cross section
is quite high. Similar arguments can easily be expanded to other backgrounds.
Others are reducible backgrounds: even though their particle contents
are similar to signal, either their cross sections are low and reconstruction
region significantly different. In that regard they satisfy only one
criterion while irreducible ones fulfill two or more.

Further we select decay channels of background events as such to give
same sign $2l^{\pm}$ with $2j$ and MET. Jets includes at least one
b-tag jet. This ensures the maximum cross section for background and
gives more contribution to histograms when we consider the detector
effects such as misidentification and over counting of particles. 

Inability to distinguish between signal and background processes increases
with misidentification of particles and loss of particles due to detector
effects. These effects causes the fuzzing of characteristics of signal
while imitating the features of signal for background processes. Moreover
b-tag efficiency plays also an important role for analyzing the signal
and background events. Since two b-tagged jets are a major property
of signal. Nevertheless two b-tagged jets requirement is so strict
for observability of signal while reducing background effects too.
Therefore we confined ourselves to at least one b-tagged jet while
recognizing characteristics of our signal and background processes.
It is also important to note that there is no interference between
signal and background at this level of calculation.

\section{Analysis}

At first stage we have started with the known limits from current
LHC experiments that put a limit on the FCNC coupling constant value
$\eta_{q}=0.07$ which is already reached and then use benchmark value
$\eta_{q}=0.07$ to insvestigate the limits for upgrading HL-LHC detector
to search for a possible FCNC signal outcome. After that we seek edge
values to limit and finalize our research. We will look forward to
push the limits for $\eta_{u+c}$, $\eta_{u}$ and $\eta_{c}$ separately.

We use the statistical significance $SS_{\mathrm{disc}}$

\begin{equation}
SS_{\mathrm{disc}}=\sqrt{2[(S+B)\ln(1+S/B)-S]}\label{eq:3}
\end{equation}
and $SS_{\mathrm{exc}}$

\begin{equation}
SS_{\mathrm{exc}}=\sqrt{2[S-B\ln(1+S/B)]}\label{eq:4}
\end{equation}
for discovery and for exclusion as given in \citep{key-22,key-23,key-24,key-25}.
For exclusion of a parameter value we are looking $SS_{\mathrm{exc}}>1.645$
corresponding to a confidence level of 95\% CL. In order to make it
complete we will give limits for discovery relation too. Both relations
reduces to $\frac{S}{\sqrt{B}}$ at large background limit. In addition,
we will conduct an evaluation in which systematic uncertainties are
estimated in order to comprehend how systematic uncertainties influence
our results. For these calculations, we will use the following formula
for discovery with systematic uncertainties

\begin{align}
SSwS_{\mathrm{disc}} & =\left[2\left((S+B)\ln\left(\frac{(S+B)(B+S^{2})}{B^{2}+(S+B)S^{2}}\right)\right.\right.\nonumber \\
 & \left.\left.-\frac{B^{2}}{\Delta_{B}^{2}}\ln\left(1+\frac{\Delta_{B}^{2}S}{B(B+\Delta_{B}^{2})}\right)\right)\right]^{\frac{1}{2}}\label{eq: 5}
\end{align}
and for exclusion case we use the equation

\begin{align}
SSwS_{\mathrm{exc}}= & \left[2\left\{ S-B\ln\left(\frac{B+S+x}{2B}\right)\right.\right.\nonumber \\
 & \left.-\frac{B^{2}}{\Delta_{B}^{2}}\ln\left(\frac{B-S-x}{2B}\right)\right\} \nonumber \\
 & \left.-\left(B+S-x\right)\left(1+\frac{B}{\Delta_{B}^{2}}\right)\right]^{1/2}\label{eq:6}
\end{align}
where $x$ being
\begin{equation}
x=\sqrt{\left(S+B\right)^{2}-\frac{4SB\Delta_{B}^{2}}{B+\Delta_{B}^{2}}}.
\end{equation}
When it comes to our analysis path, the first thing we'll do is focus
on the key points for analysis and talk about the unique characteristics
of the signaling process. These characteristics will then be disclosed
by presenting the kinematic variables, and the method to be used in
the study will be determined. Following that, the analysis will be
performed, and the results will be provided.

Here we will track exactly two positively/negatively charged leptons
as same sign lepton pairs since we investigate the case $W^{\pm}\rightarrow l^{\pm}\nu_{l^{\pm}}$
followed after $t(\bar{t})\rightarrow W^{+}b(W^{-}\bar{b})$. Missing
transverse energy is also an essential charactersitics of the process
too. We note that despite we have only two b-jets in our signal when
we consider the nature of interaction, more jets must be generated
and we need to distinguish them from bottom quarks to reconstruct
two top quarks. That point needs a little bit attention when we think
of backgrounds and to make it clear we would like to go deeper: as
we know our background events have more particles, in addition the
nature of interaction also dictates numerous jets which gives more
hadronic transverse energy. When we consider both, a cut that is limiting
the number of jets seem to be advantageous. The best choice at first
glance is limiting jet number as two, so we conclude with exact event
selection. Nonetheless, taking into account detector effects, in a
situation where two leptons are detected individually, if there are
no jets or only one jet, these jets are more likely to escape from
the detector. Working with a small number of jets is useful in this
regard, as backdrops are highly prominent when working with a large
number of jets. Furthermore, because the top quark is the source of
leptons in the processes, it can be assumed that every case in which
two same-sign leptons are seen belongs to the signal event, again
taking charge conservation into account. Again, b-tagging serves a
purpose here. Of course, without jets, this labeling is not conceivable.
However, in single-jet (or fat jet) scenarios, this criterion can
be used to provide the analysis a boost.

For lepton flavors we have 2 possibilities namely $e^{\pm}$ and $\mu^{\pm}$
for $l^{\pm}$ case since $\tau$ lepton disintegrates before reach
the detector so its analysis is out of scope. In that respect we divide
analysis region to three which includes three possibilities of same
sign lepton pairs ($e^{\pm}e^{\pm},\mu^{\pm}\mu^{\pm},e^{\pm}\mu^{\pm}$)
with exactly two jets while at least one of them b-tagged and lastly
presence missing transverse energy in events. 

Decay of top quarks in their rest frame give rise to high $p_{T}$
b-jets larger than about 80 GeV as a prediction in addition same happens
for $W^{+}$ bosons and daughter particles should have at least 40
GeV. These particles also carries momentum, thus we expect boosted
behaviour at histograms for mother and daughter particles. 

\begin{figure}[h]
\raggedright{}\includegraphics[scale=0.48]{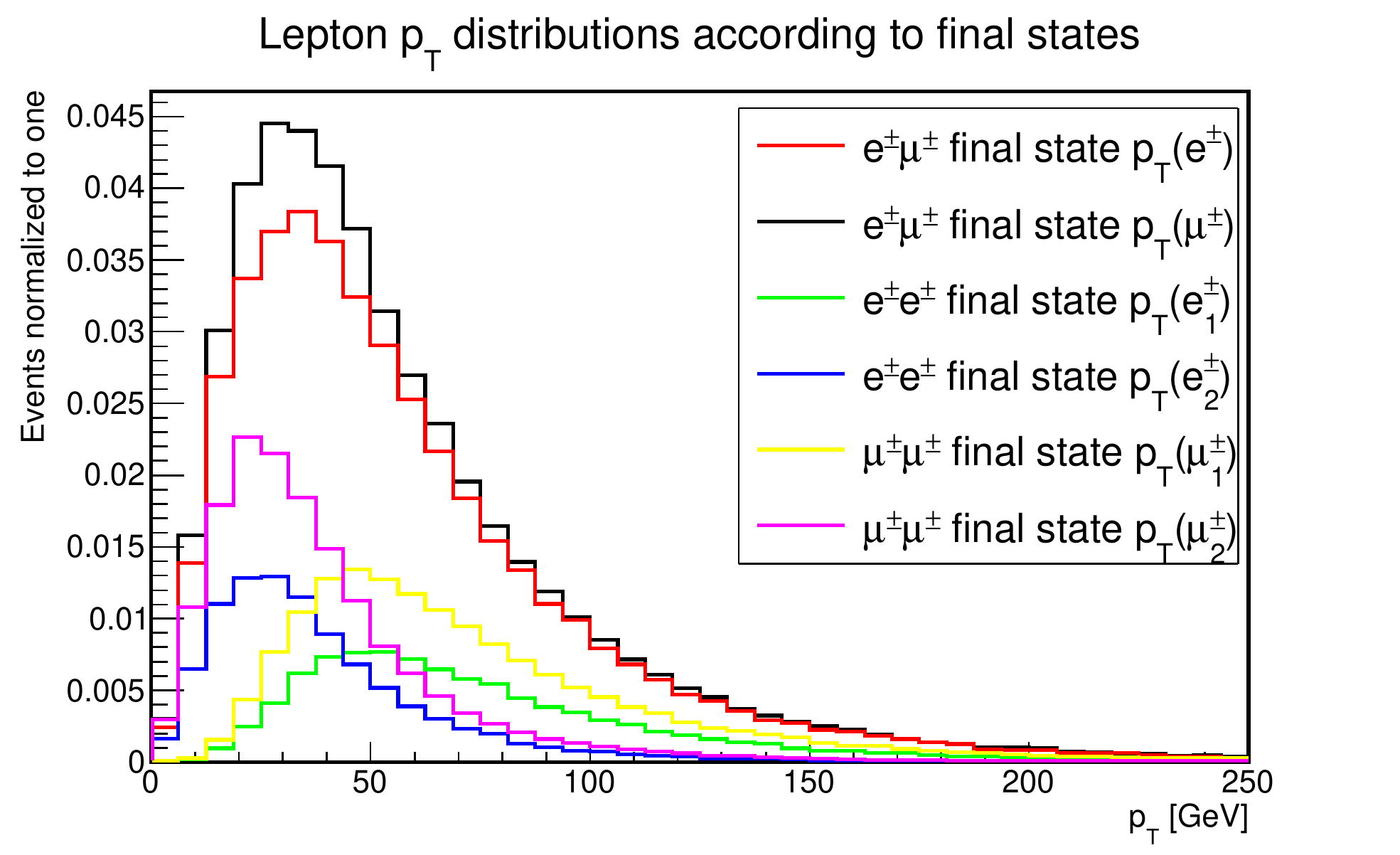}\caption{For the signal process, lepton $p_{T}$ distributions $e^{\pm}e^{\pm},\mu^{\pm}\mu^{\pm},e^{\pm}\mu^{\pm}$
event regions\label{fig:Lepton--distributions}: Histogram clearly
shows that $e^{\pm}\mu^{\pm}$ final state is more favorable. The
$e^{\pm}\mu^{\pm}$ pair comes from disintegration of $W^{\pm}$ pairs
which have about 80 GeV rest mass. Hence that energy and momentum
shared by final particles and gives a peak arround 40 GeV with boosted
behavior. However same flavor final states shows an asymmetry originates
from the following reasons: Detector always discriminates lower and
higher $p_{T}$ particle which gives a gap between first and second
highest $p_{T}$ object. Nevertheless they all have boosted behavior
and give peaks close to 40 GeV as well. }
\end{figure}

\begin{figure}
\includegraphics[scale=0.46]{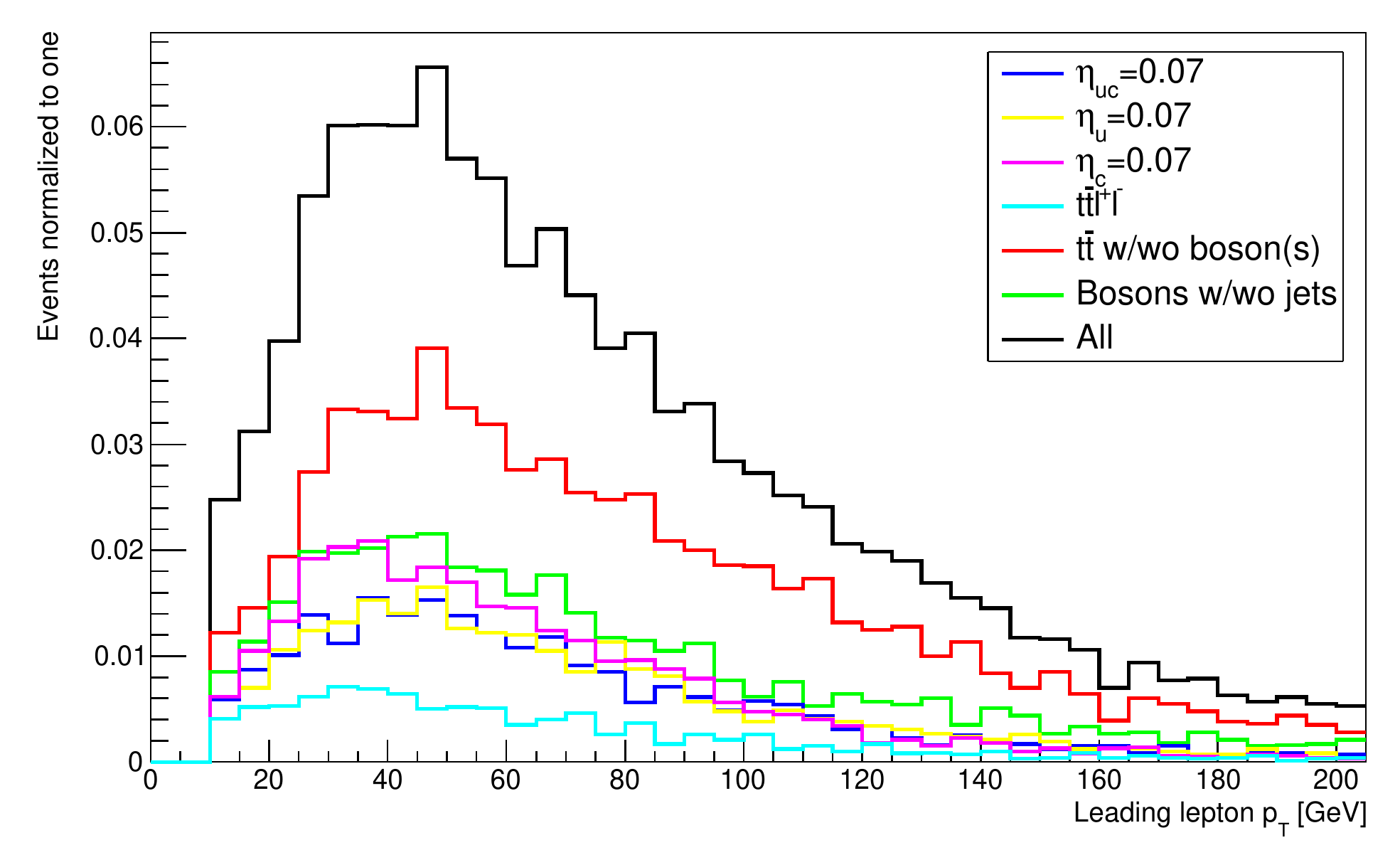}

\caption{$p_{T}$ distribution of signal and background processes for leading
leptons.\label{fig:-distribution-of}}

\end{figure}

\begin{figure}
\includegraphics[scale=0.46]{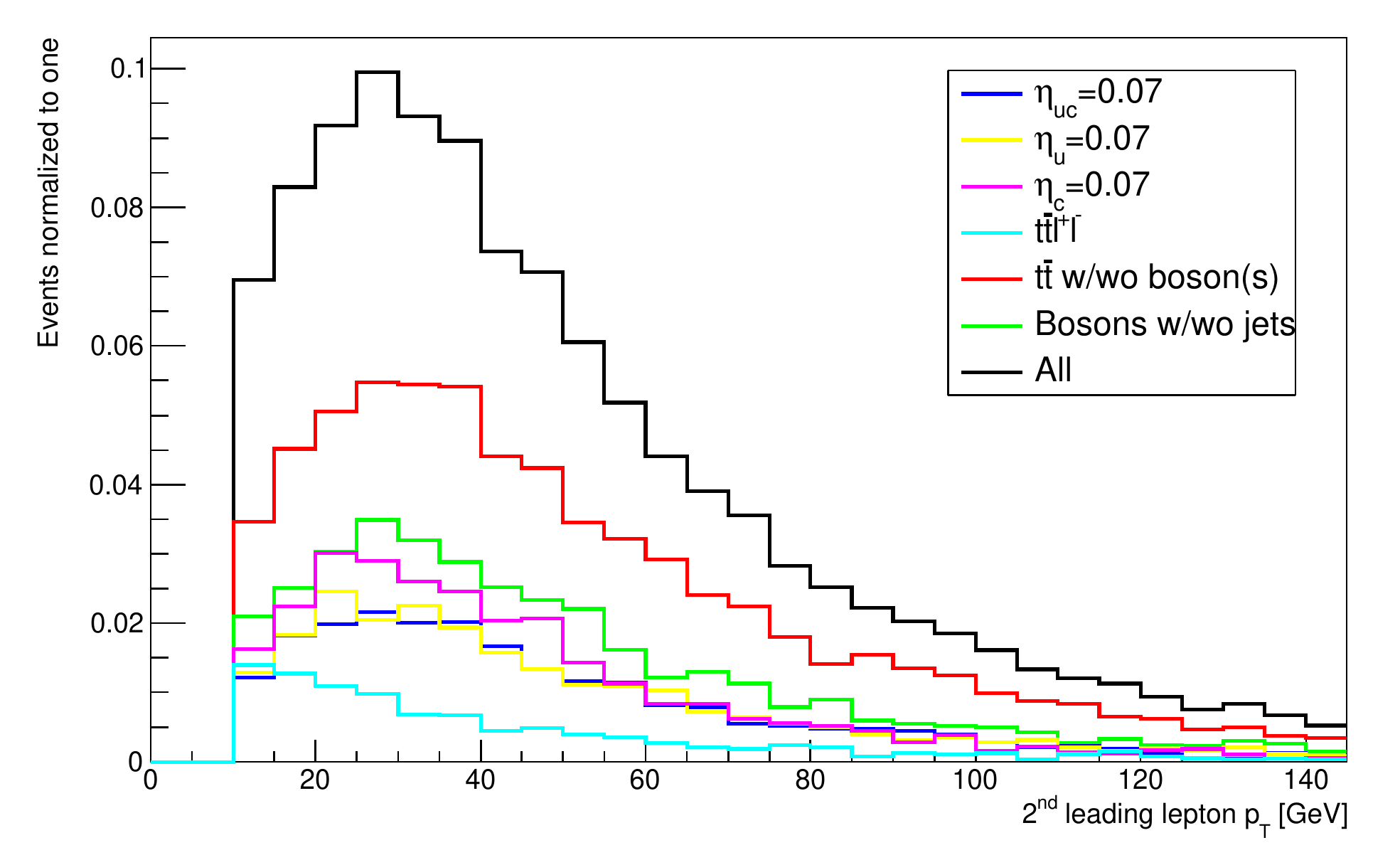}

\caption{$p_{T}$ distribution of signal and background processes for secondary
leading leptons.\label{fig:-distribution-of-1}}

\end{figure}

\begin{figure}[h]
\begin{raggedright}
\includegraphics[scale=0.46]{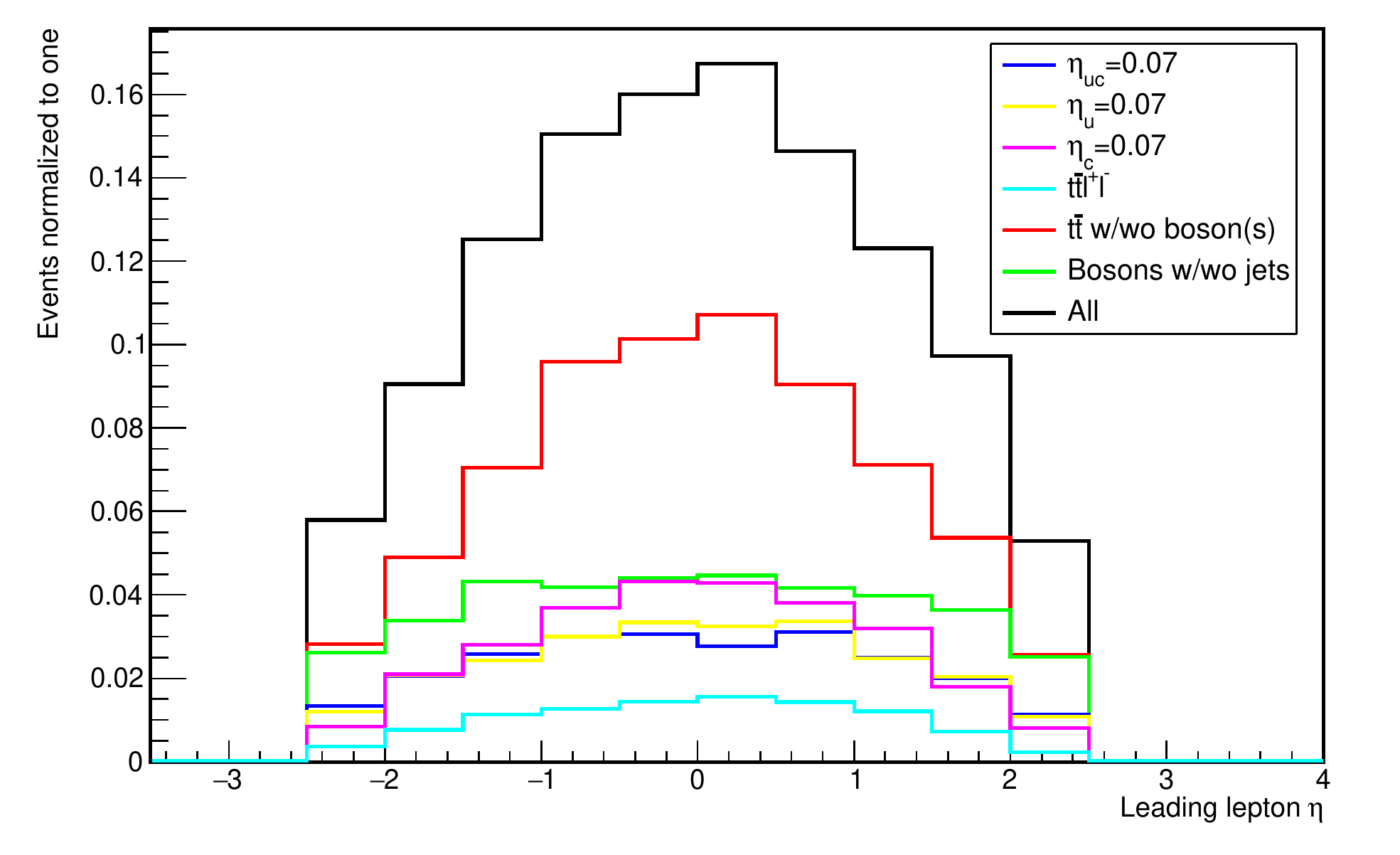}\caption{Lepton $\eta$ distributions for signal and backgrounds: general detection
is central.\label{fig:Lepton--distributionsiona}}
\par\end{raggedright}
\end{figure}

\begin{figure}
\includegraphics[scale=0.48]{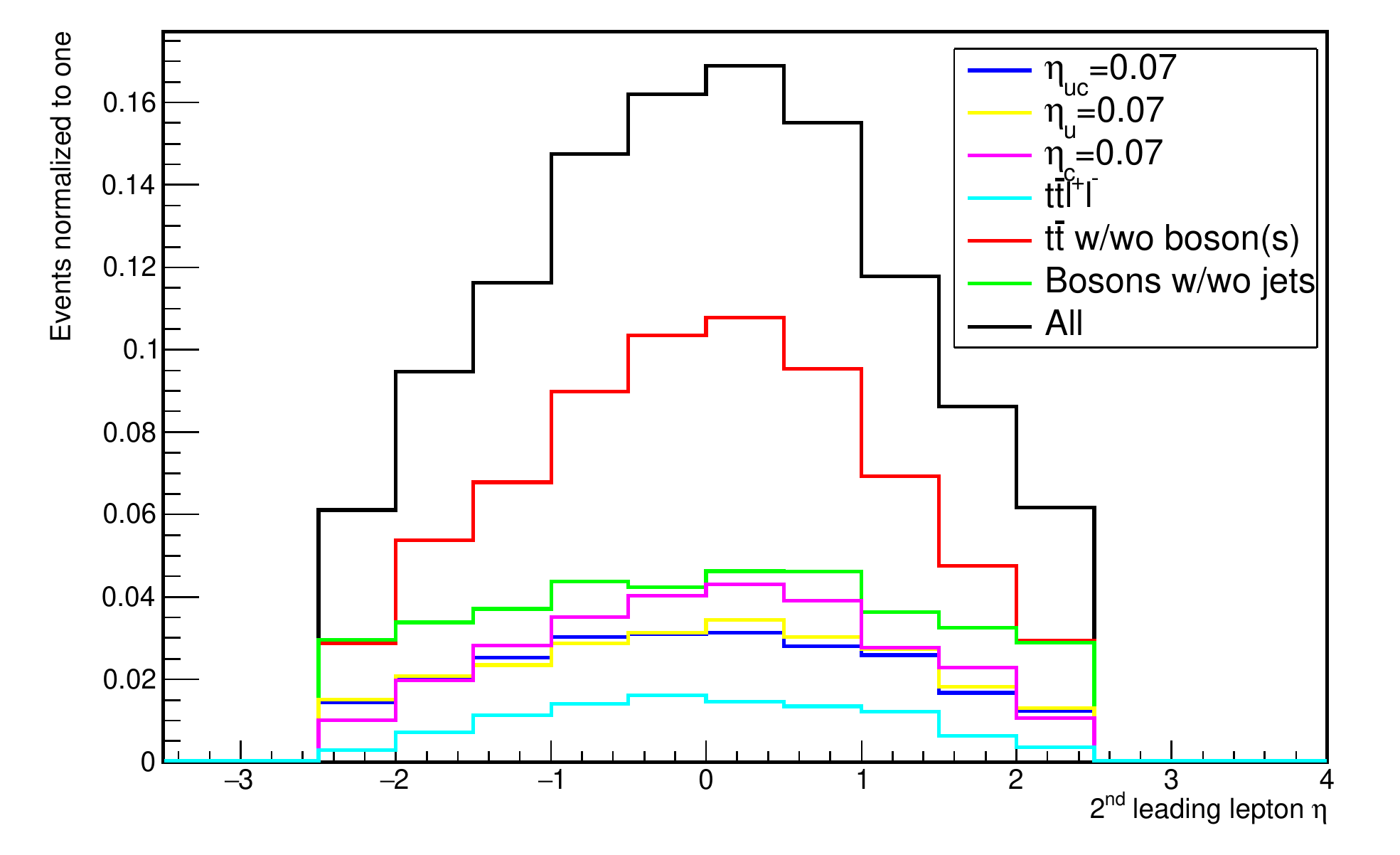}

\caption{$\eta$ distributions of the secondary lepton from signal and background.
As can be observed, they do not form a clear cut, and their significance
in the analysis is rather minor.\label{fig:-distributions-are}}

\end{figure}

\begin{figure}[h]
\begin{raggedright}
\includegraphics[scale=0.46]{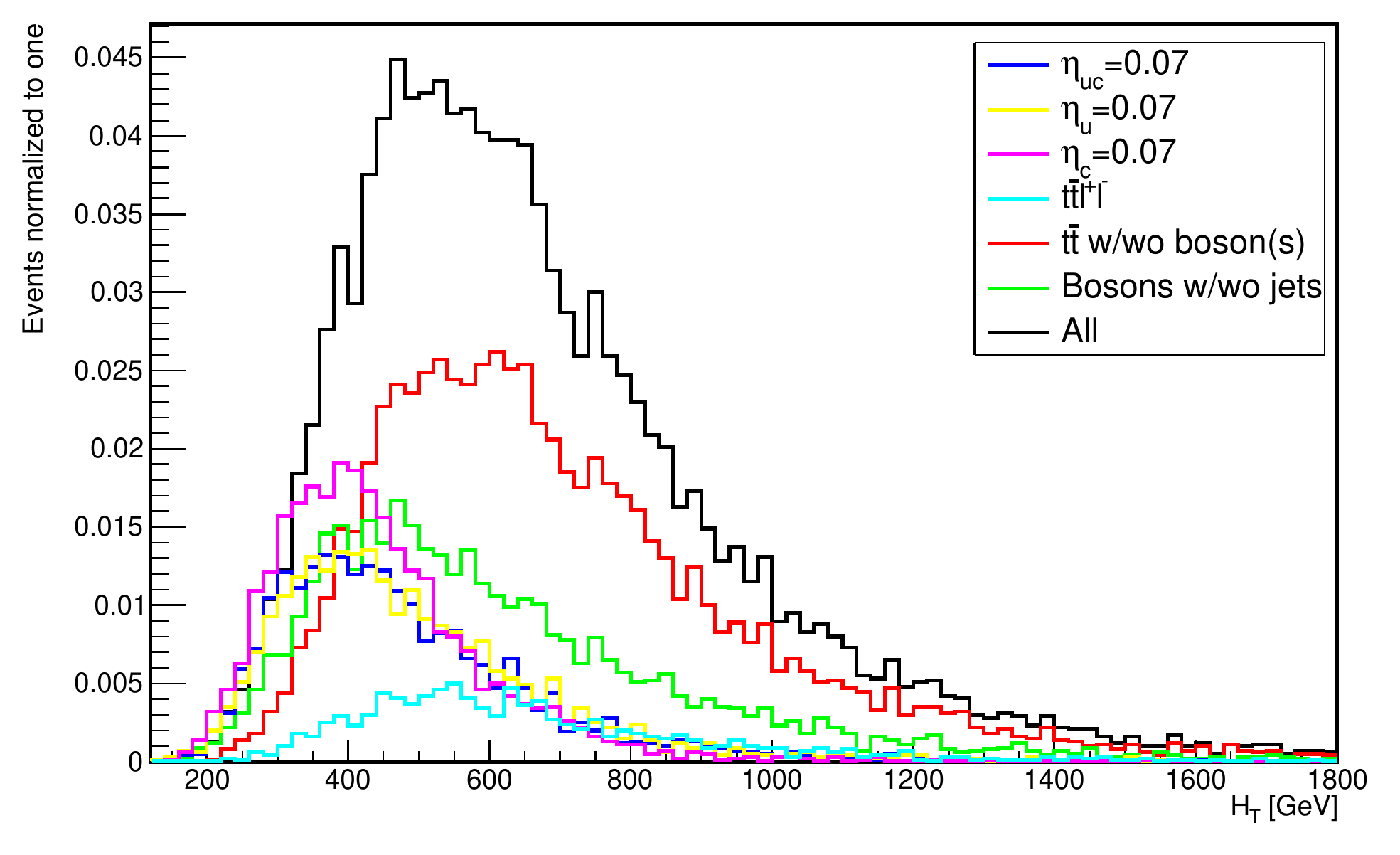}
\par\end{raggedright}
\begin{raggedright}
\caption{Scalar $H_{T}$\label{fig:Scalar-HT} distribution for the signal
and background processes. Since more jet generated at backgrounds
they have relatively shitfted to forward.}
\par\end{raggedright}
\end{figure}

\begin{figure}[h]
\includegraphics[scale=0.46]{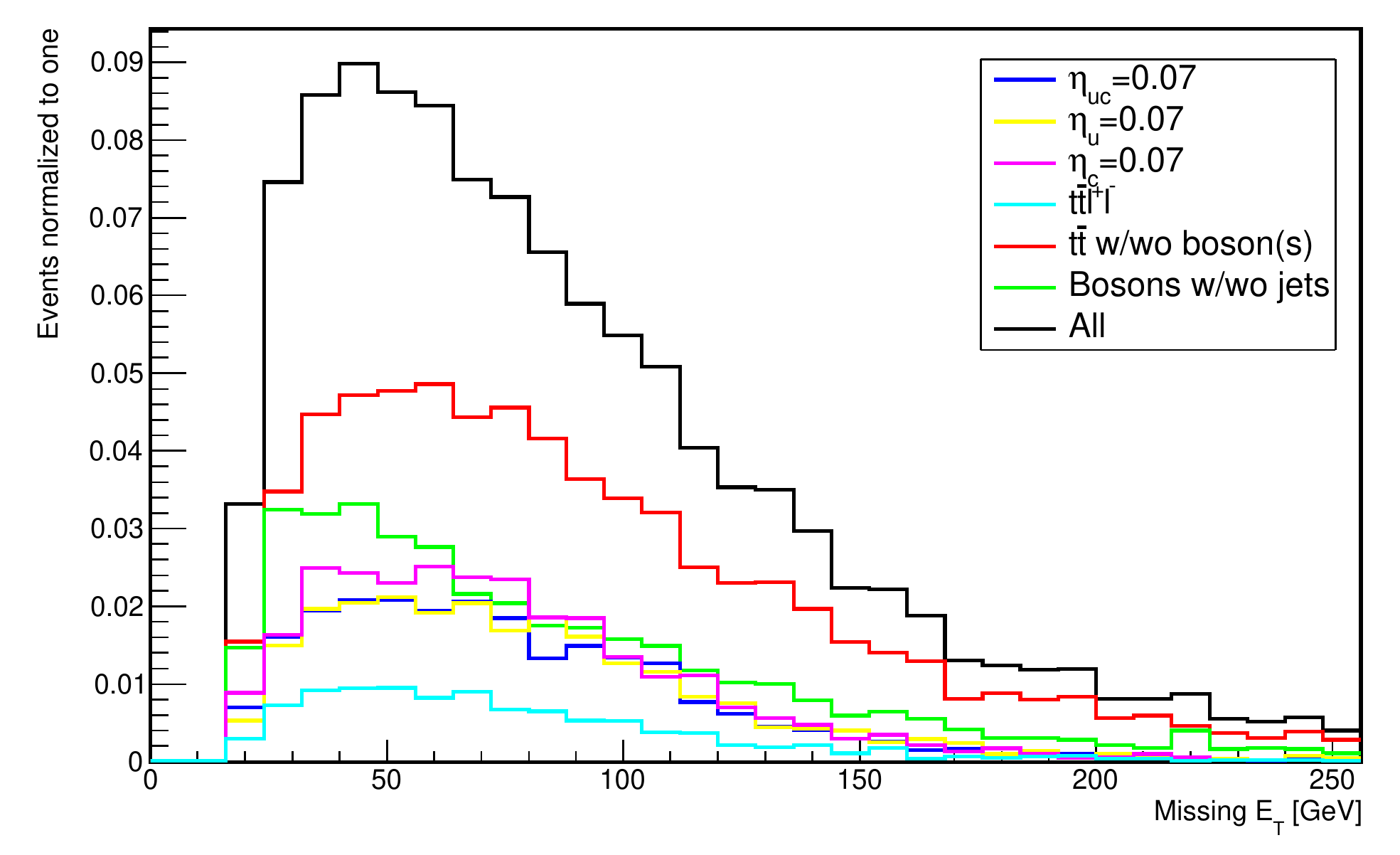}
\raggedright{}\caption{Neutrinos are the main source of missing energy of the interaction.
Here we have two neutrios coming from $W^{\pm}$ decay. Thus histogram
gives a peak about 40-50 GeV and boosted too which are in complete
consistency with our expectations.\label{fig:Neutrinos-are-the}}
\end{figure}

\begin{figure}[h]
\includegraphics[scale=0.46]{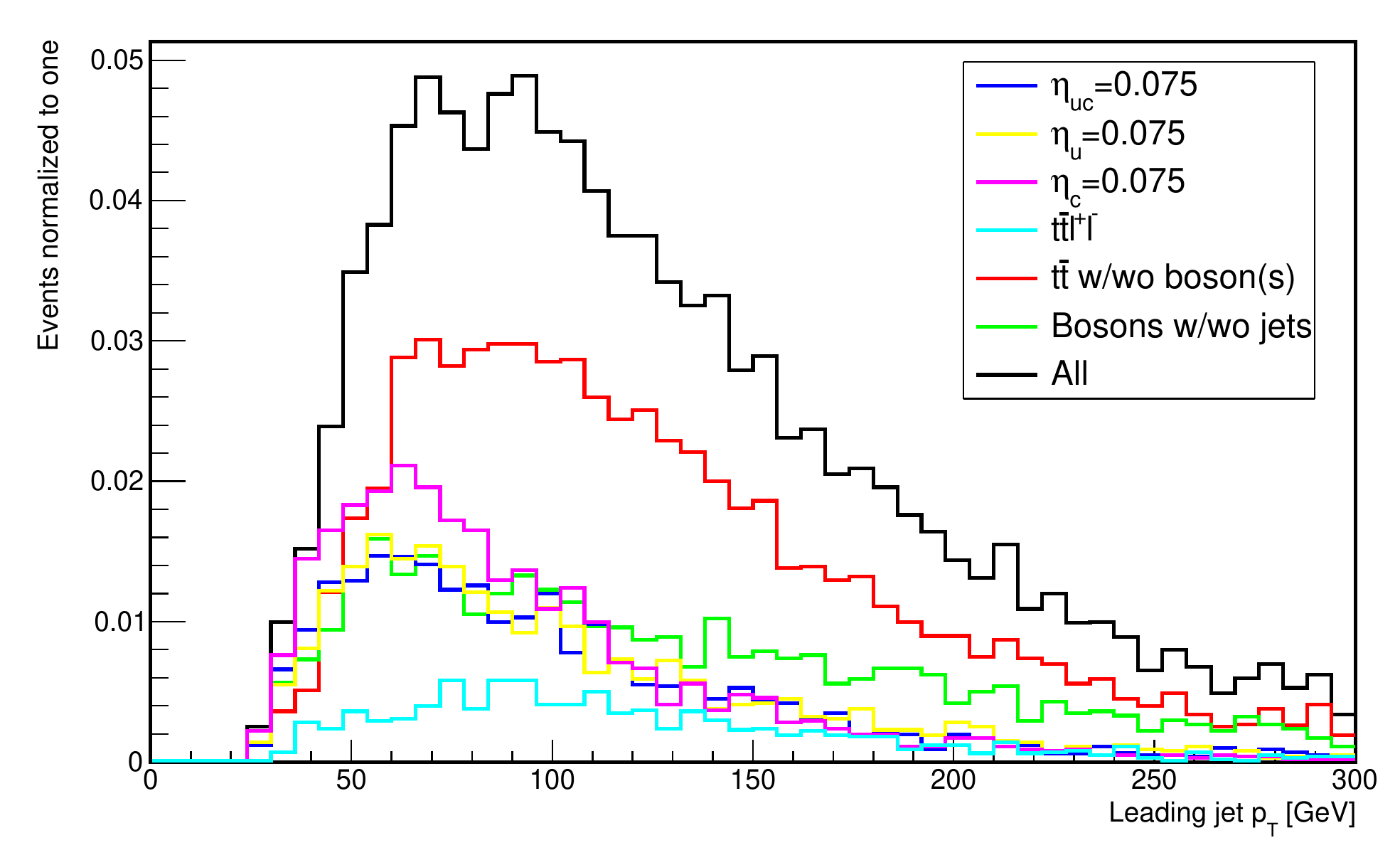}
\raggedright{}\caption{Jet $p_{T}$\label{fig:Jet}: Even though we know our process gives
mainly two b-jets, nature of process gives more jets as final state
objects. Higher number of jets lowers $p_{T}$ values for the jets
comes from leading and second leading jets and squeeze their $p_{T}$
values below 60-70 GeV. The jets playing role at reconstructing top
quarks energetic and may have $p_{T}$ above that values. Additionally,
the effect of the longitudinal component must not be overlooked.}
\end{figure}

\begin{figure}
\includegraphics[scale=0.46]{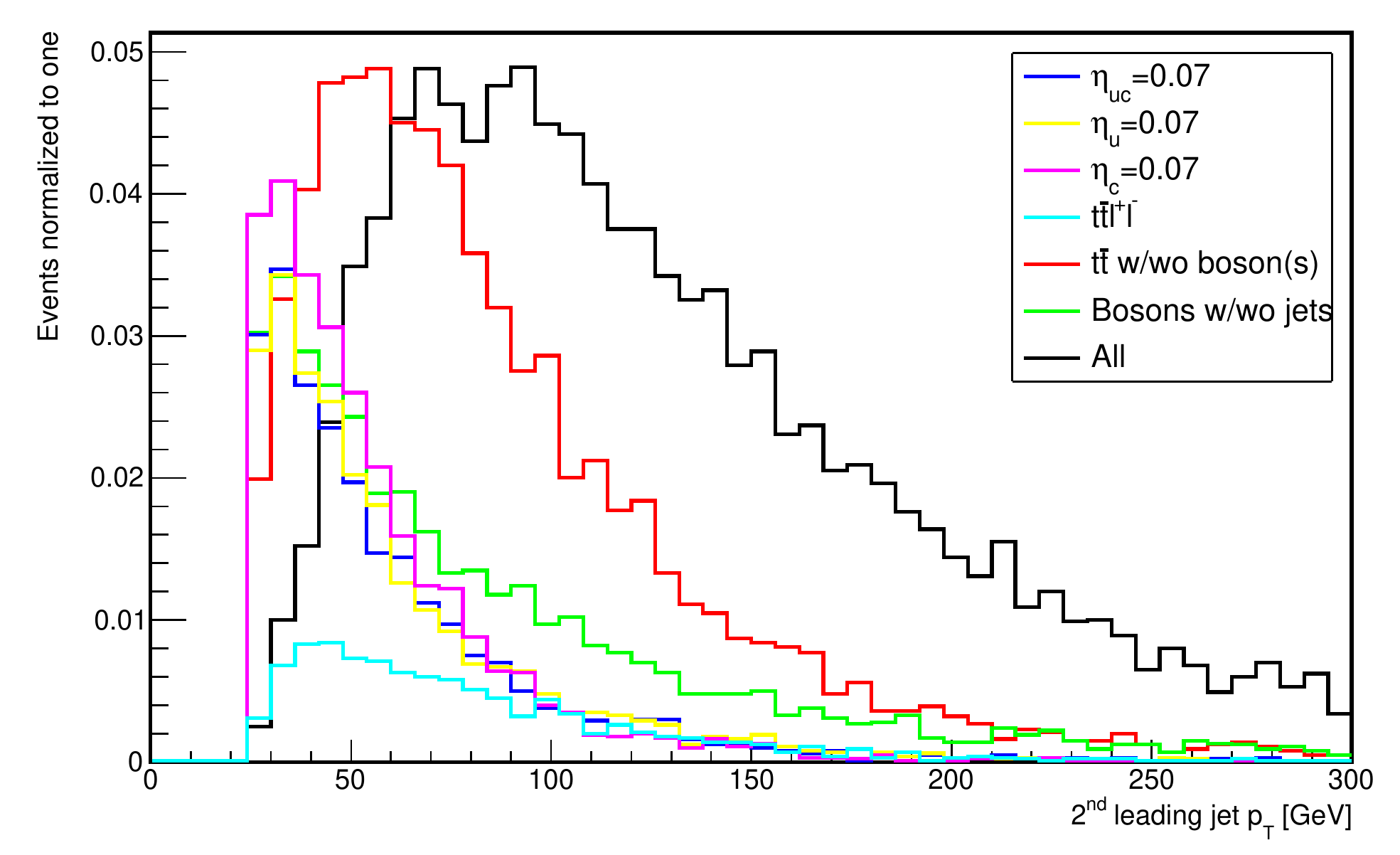}

\caption{The overall behavior of the secondary jets in the backgrounds is highly
forward, in contrast to the second jets in the signal.\label{fig:The-overall-behavior}}

\end{figure}

\begin{figure}[h]
\begin{raggedright}
\includegraphics[scale=0.46]{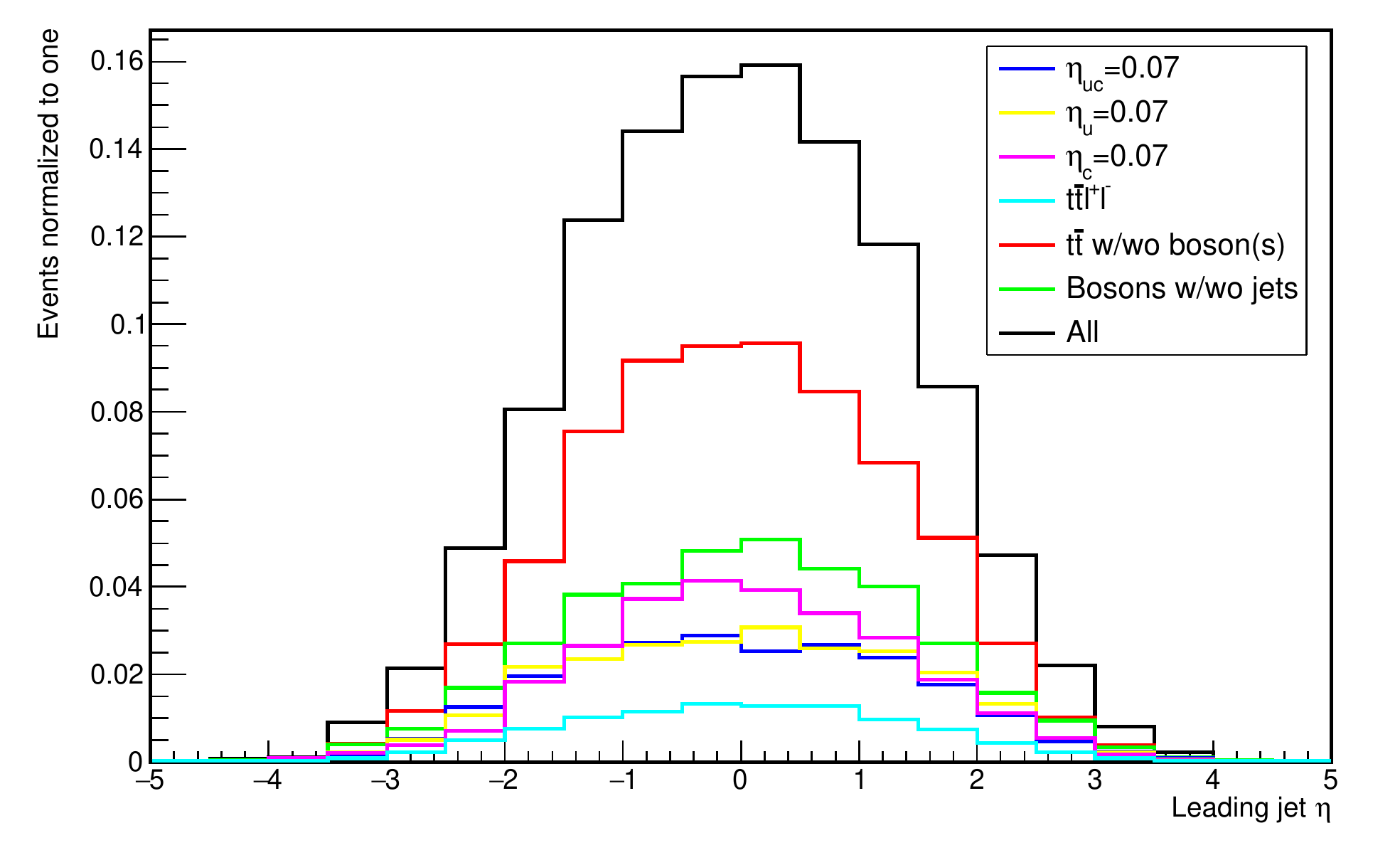}\caption{Leading jet $\eta$ distribution at detector is central.\label{fig:Jet--distribution}}
\par\end{raggedright}
\end{figure}

\begin{figure}
\includegraphics[scale=0.46]{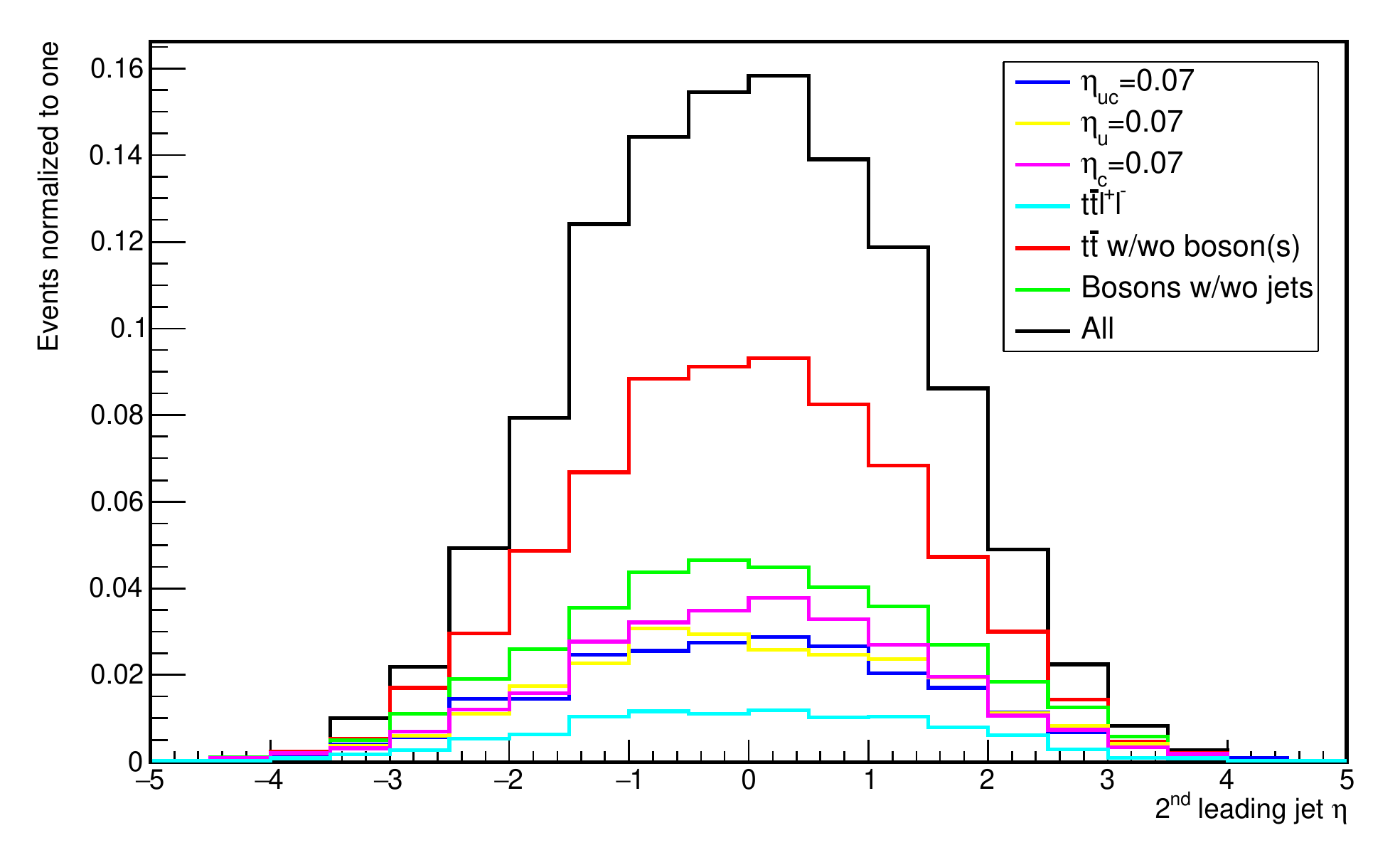}

\caption{Second leading jet $\eta$ distribution at detector is central.\label{fig:Second-leading-jet} }

\end{figure}

To sum up at the begining of the analysis we have divided signal region
to three analysis region with exact event selection, followed by simple
cuts given in Table \ref{tab:List-of-basic}.
\begin{table}[h]
\centering{}\caption{\label{tab:List-of-basic}List of basic cuts.}
\begin{tabular}{|c|}
\hline 
Event Selection and Basic Cuts\tabularnewline
\hline 
\hline 
$N(jets)\geq2$\tabularnewline
\hline 
$N(l^{\pm})=2$ (Same sign)\tabularnewline
\hline 
$p_{T}^{jets}>20\ \mathrm{GeV}$\tabularnewline
\hline 
$p_{T}^{l^{\pm}}>10\ \mathrm{GeV}$\tabularnewline
\hline 
$MET>20\ \mathrm{GeV}$\tabularnewline
\hline 
$|\eta^{l}|<2.5,\ |\eta^{j}|<5$\tabularnewline
\hline 
$\Delta R(l_{1},l_{2})>0.4$\tabularnewline
\hline 
$\Delta R(j_{1},j_{2})>0.4$\tabularnewline
\hline 
At least one b-tagged jet \tabularnewline
\hline 
\end{tabular}
\end{table}
 Here, the $\eta$ cuts were choosen to work with the more sensitive
regions of the detector for leptons especially. Furthermore, $\Delta R$
cuts were established as the minimum lepton isolation criteria. For
trigerring and good object selection criteria for jets and leptons,
missing transverse energy and $p_{T}$ cuts are minimally incorporated.
To avoid a fat-jet scenario, a $\Delta R$ cut was regarded appropriate
for jets, although jets reaching the detector were tolerated by avoiding
an $\eta$ limiting cut. As previously stated, it was noted that at
least one of the jets entering the depicted histograms was b-tagged.
Here we give the kinematical distributions for lepton $p_{T}$ at
Fig. \ref{fig:Lepton--distributions}, \ref{fig:-distribution-of}
and \ref{fig:-distribution-of-1}, for lepton $\eta$ at Fig. \ref{fig:Lepton--distributionsiona}
and \ref{fig:-distributions-are}, $H_{T}$ and MET at Fig. \ref{fig:Scalar-HT}
and \ref{fig:Neutrinos-are-the}, and lastly jet $p_{T}$ and $\eta$
at Fig. \ref{fig:Jet}, \ref{fig:The-overall-behavior}, \ref{fig:Jet--distribution},
\ref{fig:Second-leading-jet} belonging the signal process.
Finally we present histograms showing the characteristics of jets
produced and little comment on them.
\begin{figure}[h]
\raggedright{}\includegraphics[scale=0.46]{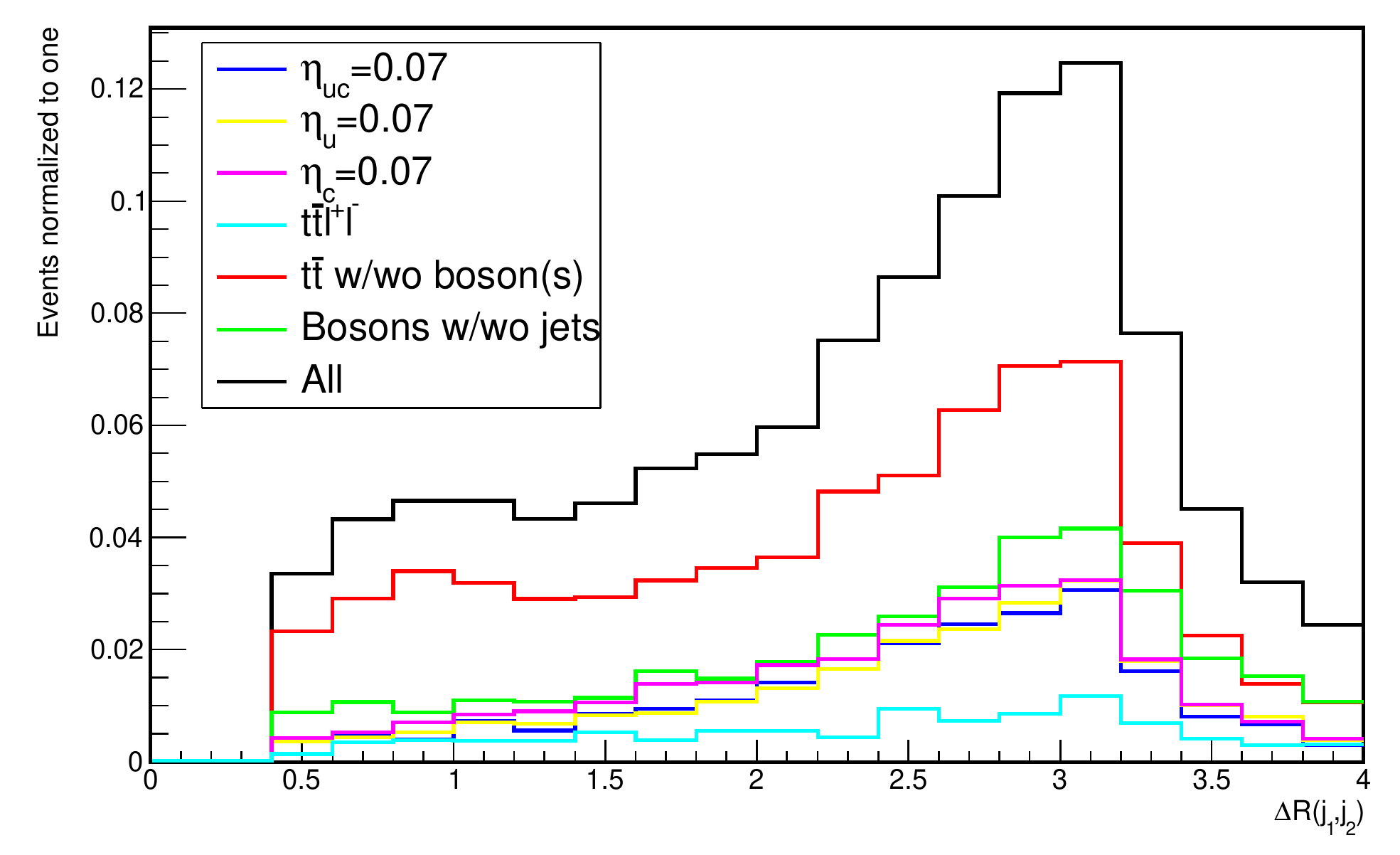}\caption{$\Delta R(j_{1},j_{2})$\label{dR j1j2} distribution between two
jets. Although there is no significant difference between the signal
and background $\Delta R$ distributions at this point, it stands
out as one of the most important variables since the signal process
is symmetrical in its stationary frame of reference. Jets have the
direct top quarks' back-to-back scattering structure.}
\end{figure}
\begin{figure}[h]

\raggedright{}\includegraphics[scale=0.46]{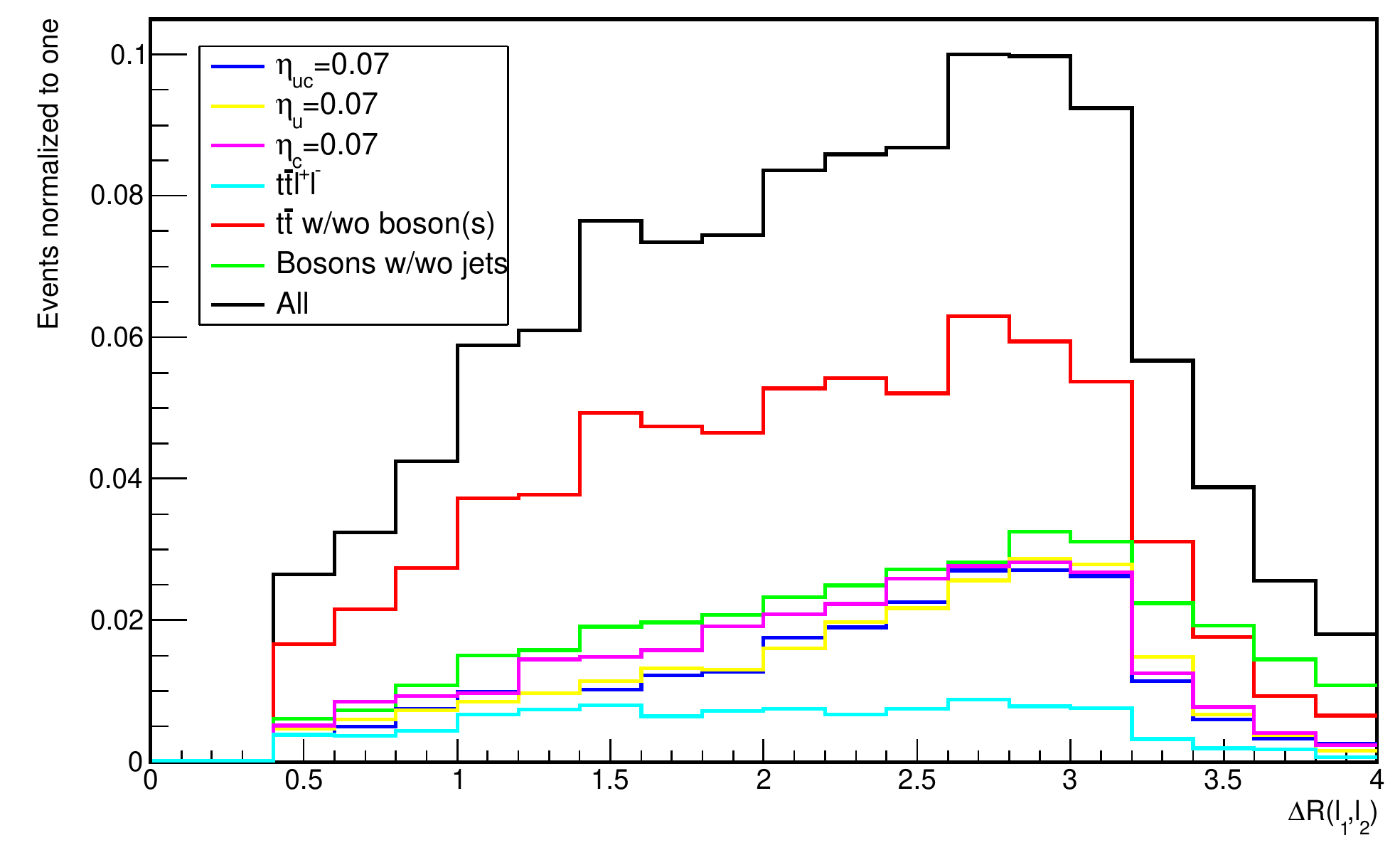}\caption{$\Delta R(l_{1}^{+},l_{2}^{+})$\label{dR l1l2} distribution between
two leptons. The $\Delta R$ variable in leptons is very crucial for
discrimination, as it is in jets, but their behavior is much looser
in comparison to jets due to additional energy-momentum conservation
constraints from the decay of the W boson.}
\end{figure}
These histograms compare the behavior of signal and background events
without delving into a detailed investigation. Except for a few cuts relevant
to the study, the segments utilized for event production have been
transferred to the detector level in order to provide the histograms
in their simplest form. While some variables reflect differences.
Although it is possible to separate the signal process, which includes
two new physics vertices (and lowers cross section drastically). In
the background, due to the high cross-section of the $t\bar{t}W^{\pm}$
background and its similarity to the signal, it is not possible to
make a discrepancy after a point and provide the desired improvement
in the analysis.

Although this makes the investigated process more appealing for exclusion,
because the path forward with cut-based analysis is limited, better
results can be obtained by utilizing machine learning techniques with
the help of variables defined after these basic cuts.

The most fundamental variables in this analysis are the $p_{T}$,
$\eta$ and $\phi$ components of the jets up to 4th as well as the
same kinematic variables as the first two leptons. Furthermore, variables
such as missing $E_{T}$, $H_{T}$, and $\Delta R$ are used together
with, the invariant masses of the two jets and two leptons, the invariant
and transverse masses in the quadruple state $(l_{1},l_{2},j_{1},j_{2})$,
and finally $m_{T}^{W_{1,2}}$, $m_{T}^{t_{1,2}}$ reconstructions
for the final state particles which are important for the result.
For $m_{T}^{W}$ and $m_{T}^{t}$ variables we have used the relations,
\begin{align}
m_{T}^{t} & =\left[\left(\sqrt{\left(p^{l}+p^{b}\right)^{2}+\left|\vec{p}_{T}^{l}+\vec{p}_{T}^{b}\right|^{2}}+\left|\vec{p}_{T}^{\nu_{l}}\right|\right)^{2}\right.\nonumber \\
 & -\left.\left|\vec{p}_{T}^{l}+\vec{p}_{T}^{b}+\vec{p}_{T}^{\nu_{l}}\right|^{2}\right]^{1/2}
\end{align}
and 
\begin{equation}
m_{T}^{W}=\sqrt{2p_{T}^{l}E_{T}^{\mathrm{miss}}-\vec{p}_{T}^{l}.\vec{p}_{T}^{\nu_{l}}}.
\end{equation}

During the generation of these variables, cuts identical to those
in Table \ref{tab:List-of-basic} were utilized, with minor modifications.
First, the event selection regions are separated immediately into
two lepton regions with same signs, regardless of the number of jets.
While the fundamental sections of the $p_{T}$ cuts were kept, the
criteria for lepton and jet separation were abandoned. In addition,
the $p_{T}$ cuts for the fifth jet remained at 15 GeV. Each $\eta$
variable is set to a value less than 2.5. To contrast the prominent
signal regions in low jet number and low $H_{T}$ states with the
dominant background processes in high jet number and high $H_{T}$
states, the number of jets is included up to a maximum of five and
processes begin in non-jet states. In addition, the parsing of the
$\Delta R$ variable was to be performed solely using machine learning
techniques\citep{key-47}.

\begin{figure}
\includegraphics[scale=0.33]{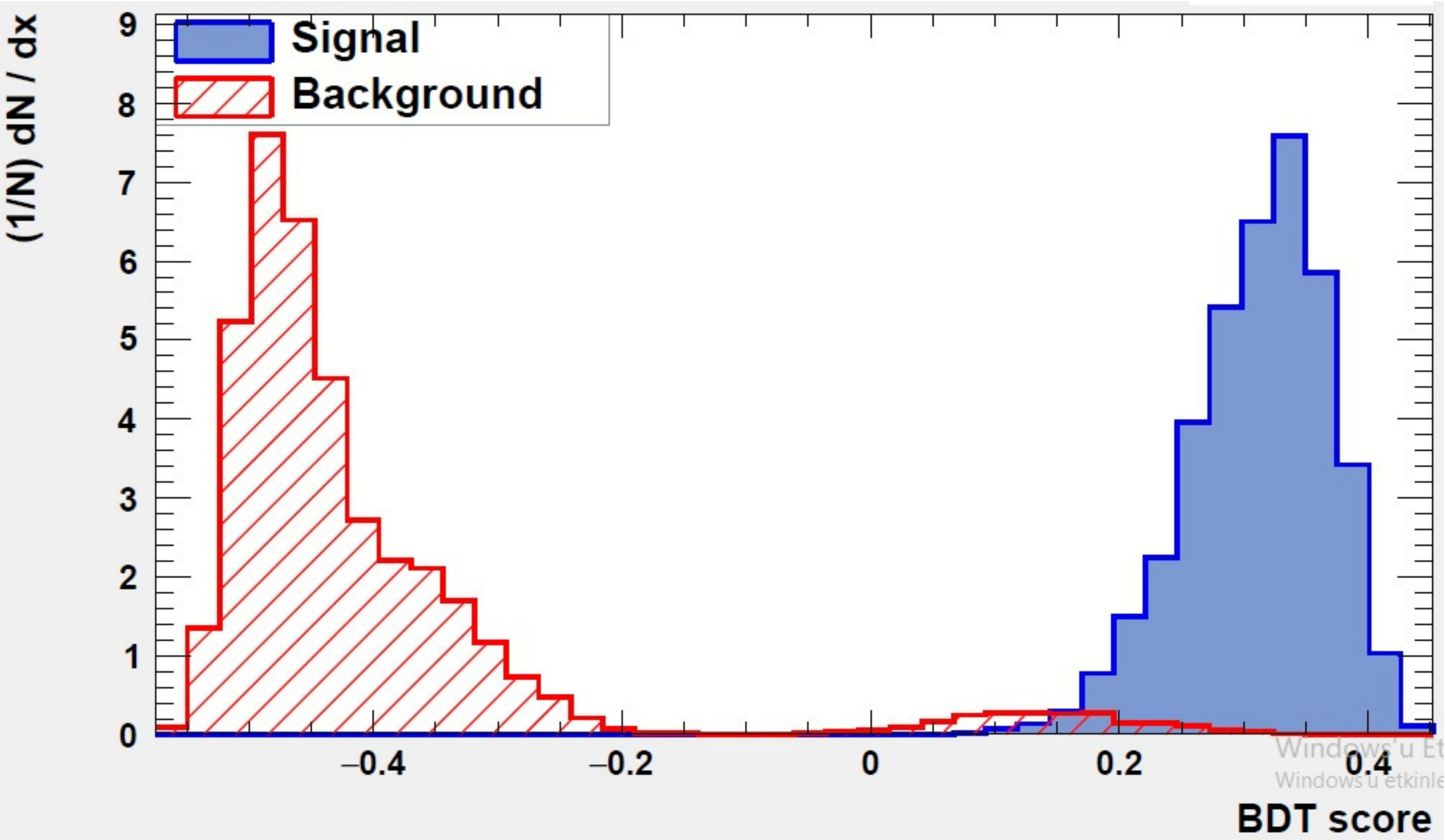}

\caption{By employing an appropriate cut, it is achievable to distinguish the
signal from the background with great efficiency. (The curve was determined
by analyzing $\eta_{uc}=0.07$.)\label{fig:By-employing-an}}
\end{figure}

Finally, the findings of the BDT analysis are presented in Fig. \ref{fig:By-employing-an},
\ref{fig:In-the-analysis,}. Observing the nonlinear behavior of the
signal and background, suitable approaches were chosen. Since a method
based on fluctuations, such as BDT, is employed. A decision tree takes
a set of input features and splits input data recursively based on
these features. Boosting is a method of conbining many weak learnings
(trees) into a strong classifier. It has been confirmed that the rate
of discriminating gradually increases between the training and testing
phases. Fig. \ref{fig:By-employing-an} demonstrates that, as a result
of the employment of several variables with high event numbers, the
distribution and height of the signal's curve are significantly superior
to the back one.
\begin{figure}
\includegraphics[scale=0.46]{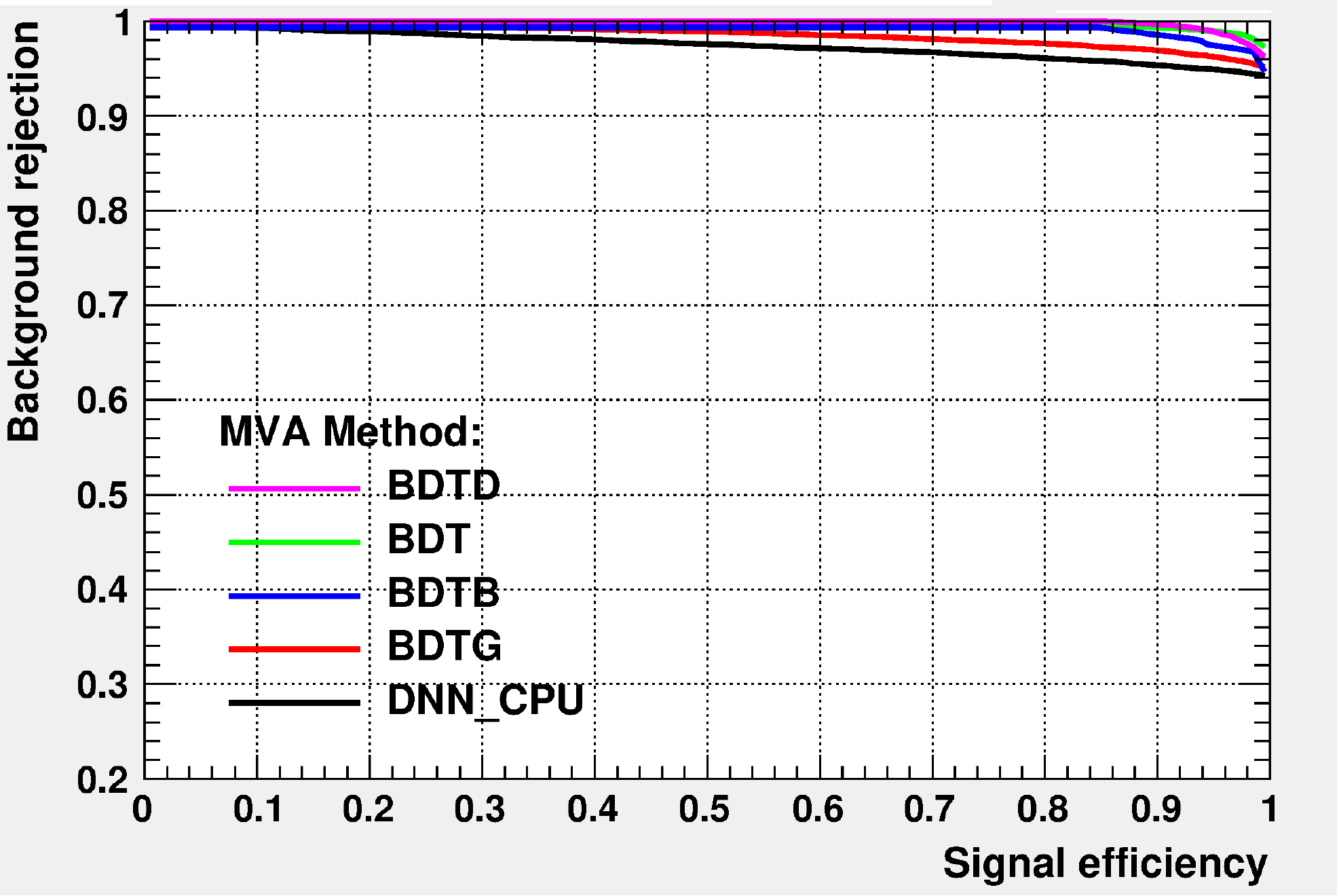}

\caption{In the analysis, nonlinear approaches were predominantly employed.
This is due to the fact that the structure of the signal and background
distributions is more accurately reflected in this manner. In this
regard, linear approaches such as Fisher's and its derivatives are
unsuitable for analysis. Again, similar procedures were not adopted
since they did not produce successful results. However, other nonlinear
approaches were incorporated in this context so that the analysis
could be compared to other ways and its evolution could be observed.
The results indicate that the inputs are uniformly distributed, no
overtraining was seen, and the analysis produced a high level of diversification
overall.\label{fig:In-the-analysis,}}

\end{figure}

\section{Results and Conclus\i ons }

\begin{figure}[h]
\includegraphics[scale=0.48]{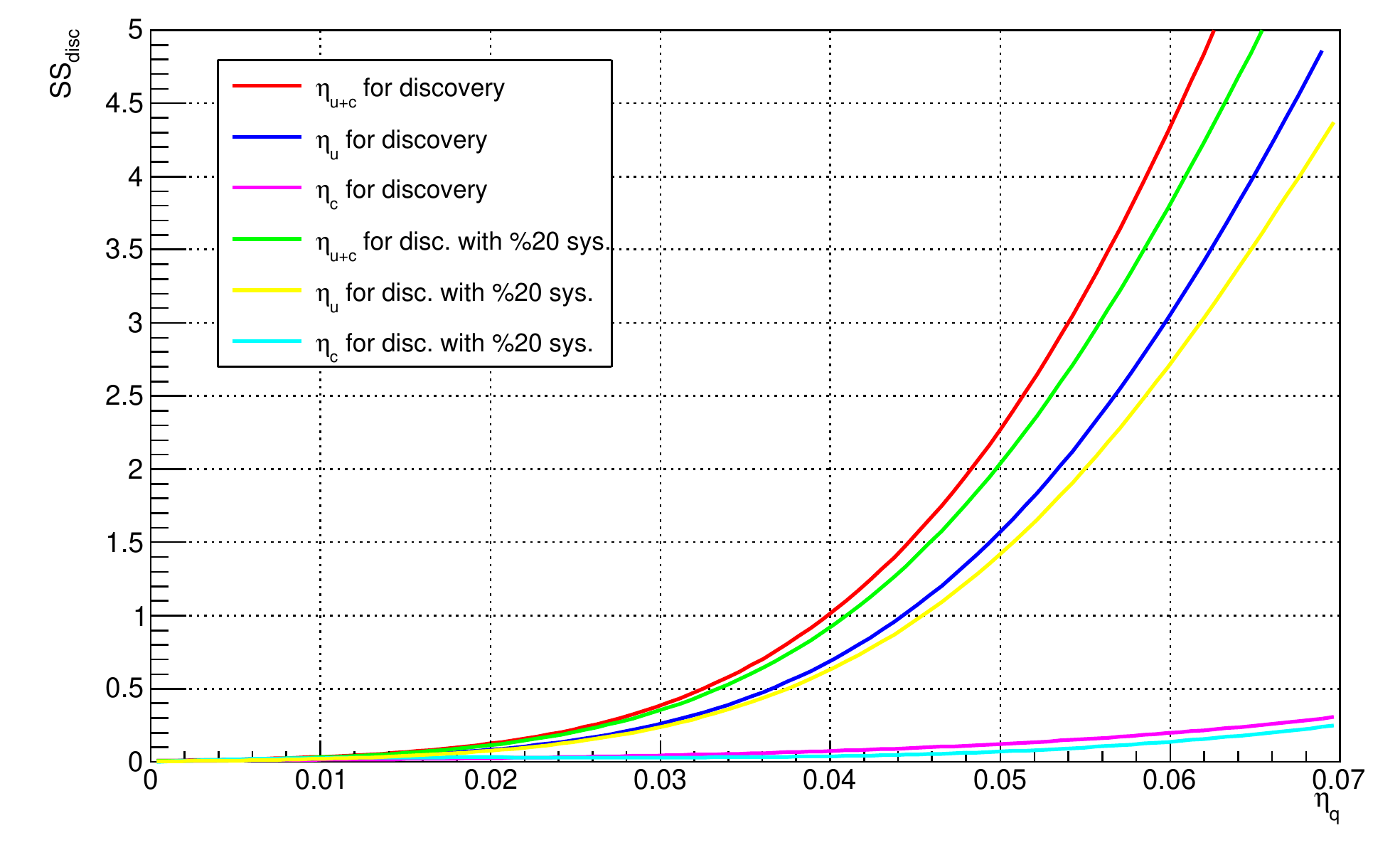}
\raggedright{}\caption{Signal significance ($SS_{\mathrm{disc}}$) versus $\eta_{q}$ coupling
parameter for three different scenarios at $3\ ab{}^{-1}$ integrated
luminosity.\label{fig:Signal-significance-versus}}
\end{figure}

\begin{figure}
\includegraphics[scale=0.48]{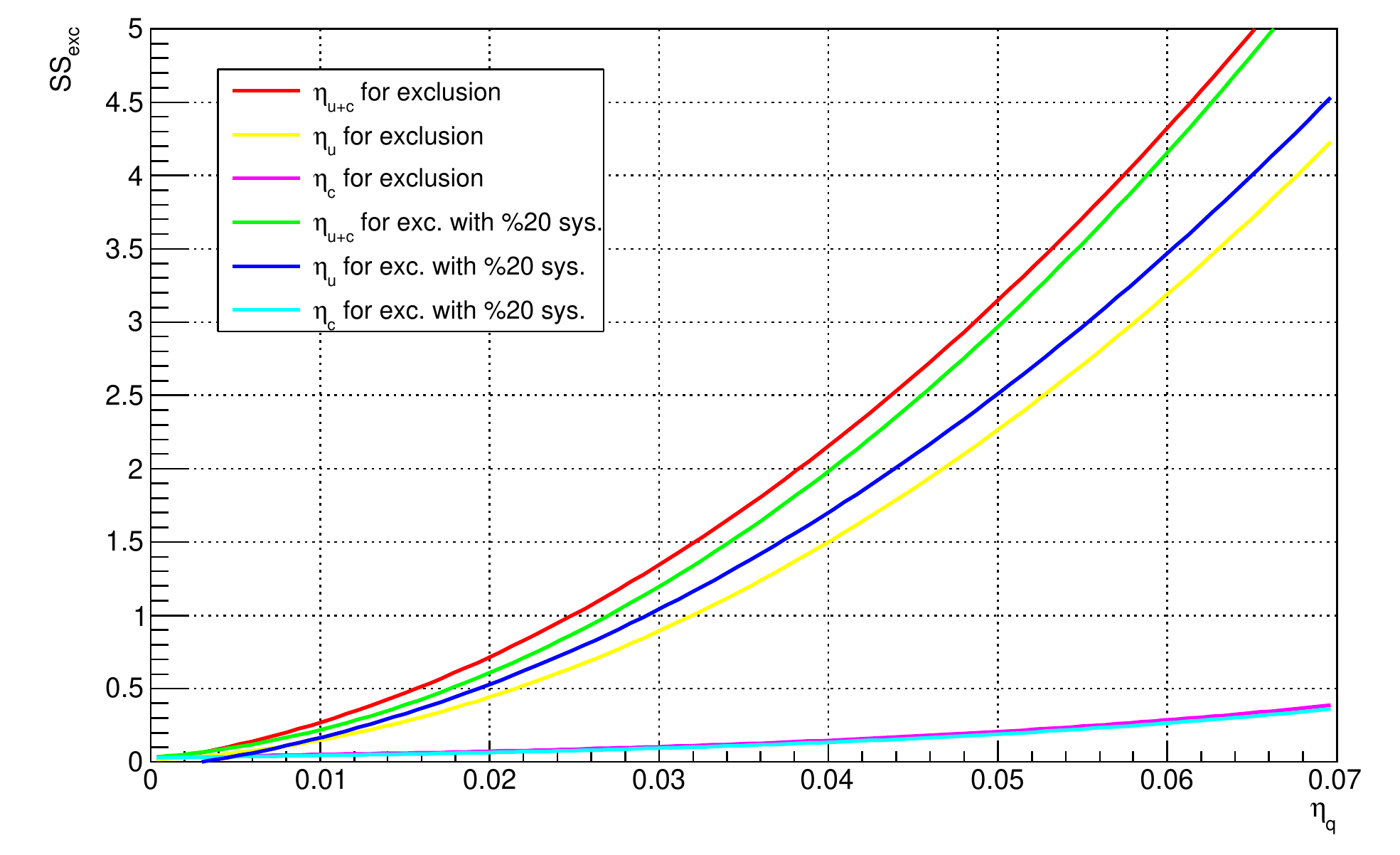}

\caption{Signal significance ($SS_{\mathrm{exc}}$) versus $\eta_{q}$ coupling
parameter for three different scenarios at $3\ ab{}^{-1}$ integrated
luminosity.\label{fig:Signal-significance-()}}

\end{figure}

In this study we have searched for accesible limits for top-Higgs
FCNC couplings using same sign lepton channel at the HL-LHC (see Fig.
\ref{fig:Signal-significance-versus}, \ref{fig:Signal-significance-()}).
This channel gives clean signal signature in addition to its low reducible/irreducible
background. However, this channel suffers from two new physics vertices.
Thus, these effects lower the cross section drastically which is a
disadvantageous feature of this analysis. Keeping these in mind we
can conclude that: simulation of this process with the same sign lepton
channel turns into a laboratory for testing the mentioned scenarios
in the text. In this respect this channel determines the upper limit
for couplings and benefits exclusion limits rather than discovery.

We have started with coupling constant $\eta_{q}=0.07$ to demonstrate
the characteristics of signal an catch the limits given in Ref. \citep{key-5}
whose limits are more or less same as our benchmark value. Then, as
stated at introduction section we have tried to improve our results
and get better limits for FCNC couplings. 

In Table \ref{tab:Limitations-on-} we summarize our analysis results
with the discovery and exclusion significance. Due to the sensitivity
of the study to exclusion, examining exclusion instances first will
expose the results more clearly. First, the results of the initial
review have improved the known limits \citep{key-5,key-6,key-92},
with the exception of some channels \citep{key-92}. In addition,
although how the results will be compared with one another is discussed
in the second chapter, these changes will be discussed in greater
detail here.

As expected, the $\eta_{u+c}$ instance produced the best results.
Cases $\eta_{u}$ and $\eta_{c}$ followed these outcomes, respectively.
Theoretically and empirically, the results vary little in the idealized
scenario, assuming a total of 20\% systematic uncertainty. This means
that the background in the study has been eliminated with great success,
and these uncertainties will not significantly impact the outcomes.
In the case of exclusion, we have improved the coupling constant limits
for $\eta_{u+c}$ and $\eta_{u}$ situations, although our limits
for $\eta_{c}$ are more stringent. However, the branching ratios
for the $\eta_{c}$ situation appear to have already been exceeded.
Similarly, the $\eta_{u}$ and $\eta_{u+c}$ scenario exceeds the
LHC constraints by a small amount. At this point, as the effective
Lagrangiande utilized by CMS contains a weak interaction constant,
it is apparent that the results may stray further from the known limitations
with this factor, despite the fact that the analysis indicates the
reverse. From the obtained coupling constants, the resulting branching
ratios are determined. The minimum values coupling constants can attain
are proportional to the number of events in the analysis, or indeed
the cross section. When the analyzed process consists of two vertices,
it is dependent on constants of the fourth order and has an advantage
proportionate to the inverse square of the coupling constant size
when compared directly. In fact, this circumstance nullifies the influence
of the weak interaction constant and drastically decreases the results
below the known levels. In this regard, it becomes evident why the
channel is superior for exclusion and why it establishes very strict
upper limits. Nonetheless, these constraints also limit other studies
of top quark-Higgs FCNC interactions. Since these restrictions are
precluded for this channel with two vertices, it stands to reason
that studies with a single vertex will go below this limit, at least
proportional to the obtained coupling constant value. Note that, we
have also caught the phenomenelogical limits for HL-LHC expected \citep{key-10,key-11,key-12,key-13,key-40}.
At that studies $\eta_{q}$ varies near 0.04 (Bear in mind that models
does not include  additional $\frac{1}{\sqrt{2}}$ factor \citep{key-11,key-12,key-13}).
\footnote{However, we would like to point out that some of these research focus
on luminosity variation rather than limit values.} 

Concerning the scenario of discovery, the limits reached for discovery
coincide with the limits reached by the CMS and ATLAS collaborations
at lower total luminosity values. In this instance, if further data
is obtained, this value indicates that exploration is feasible up
to the region's limit. However, it should not be forgotten that due
to the nature of the analysis channel, they are still upper limitations.
In this context, it has been proved that the values for an analysis
involving a single FCNC vertex can be reduced.

Lastly upcoming colliders will provide better visions for FCNC interactions
\citep{key-9}. To compare them with each other, we may say HL-LHC
and FCC-eh are expected to work at same region. Moreover, HL-LHC offers
better limits when we compare it with ILC/CLIC \citep{key-9}. So
our results have some implications on the analysis have been done
for both FCC-eh and ILC/CLIC. In support of this, studies gives similar
results to ours done for FCC-eh \citep{key-91}. Even though results
of HL-LHC will give direction to new researches without any doubt,
there is a gap in COM and luminosity values between HL-LHC and FCC-hh.
It is expected to taken down limits even further by FCC-hh. FCC-hh
can possibly rule out RS models, and start to penetrate the MSSM region.

To sum up our findings we may say, while these limits are compatible
with the expectations from HL-LHC, which enforce limitations for findings
on other channels: Since on the one hand this channel gives its clean
signal fingerprint, on the other hand even lower cross section, the
same-sign lepton channel provides upper limits and provides hints
to other detectors and thanks to its clean signal fingerprint, it
also imposes partial limitations on other channels. Our limits can
also be combined with the other sensitive channels for similar scenarios.
\begin{table}
\begin{raggedright}
\caption{Upper limits on $\eta_{q}$ parameter and corresponding branching
ratio as a potential discovery scenario projection at HL-LHC with
no systematics (left side) and with 20\% systematics (right side)
at $3\ ab{}^{-1}$ integrated luminosity.\label{tab:Limitations-on-}}
\begin{tabular}{|c|c|c|}
\hline 
Scenario & \multicolumn{2}{c|}{$SS_{\mathrm{disc}}\geq2$}\tabularnewline
\hline 
\hline 
$\eta_{u}=\eta_{c}$ & $0.048$ & $6.00\times10^{-4}$\tabularnewline
\hline 
Only $\eta_{u}$ & $0.053$ & $7.32\times10^{-4}$\tabularnewline
\hline 
Only $\eta_{c}$ & $0.12$ & $3.74\times10^{-3}$\tabularnewline
\hline 
\end{tabular}%
\begin{tabular}{|c|c|c|}
\hline 
20\% sys. & \multicolumn{2}{c|}{$SS_{\mathrm{disc}}\geq2$}\tabularnewline
\hline 
\hline 
$\eta_{u}=\eta_{c}$ & $0.050$ & $6.52\times10^{-4}$\tabularnewline
\hline 
Only $\eta_{u}$ & $0.055$ & $7.88\times10^{-4}$\tabularnewline
\hline 
Only $\eta_{c}$ & $0.12$ & $3.74\times10^{-3}$\tabularnewline
\hline 
\end{tabular}
\par\end{raggedright}
\begin{raggedright}
\begin{tabular}{|c|c|c|}
\hline 
Scenario & \multicolumn{2}{c|}{$SS_{\mathrm{disc}}\geq3$}\tabularnewline
\hline 
\hline 
$\eta_{u}=\eta_{c}$ & $0.054$ & $7.60\times10^{-4}$\tabularnewline
\hline 
Only $\eta_{u}$ & $0.060$ & $9.38\times10^{-4}$\tabularnewline
\hline 
Only $\eta_{c}$ & $0.13$ & $4.39\times10^{-3}$\tabularnewline
\hline 
\end{tabular}%
\begin{tabular}{|c|c|c|}
\hline 
20\% sys. & \multicolumn{2}{c|}{$SS_{\mathrm{disc}}\geq3$}\tabularnewline
\hline 
\hline 
$\eta_{u}=\eta_{c}$ & $0.056$ & $8.17\times10^{-4}$\tabularnewline
\hline 
Only $\eta_{u}$ & $0.062$ & $1.00\times10^{-3}$\tabularnewline
\hline 
Only $\eta_{c}$ & $0.14$ & $5.09\times10^{-3}$\tabularnewline
\hline 
\end{tabular}
\par\end{raggedright}
\begin{raggedright}
\begin{tabular}{|c|c|c|}
\hline 
Scenario & \multicolumn{2}{c|}{$SS_{\mathrm{disc}}\geq5$}\tabularnewline
\hline 
\hline 
$\eta_{u}=\eta_{c}$ & $0.063$ & $1.00\times10^{-3}$\tabularnewline
\hline 
Only $\eta_{u}$ & $0.070$ & $1.28\times10^{-3}$\tabularnewline
\hline 
Only $\eta_{c}$ & $0.15$ & $5.83\times10^{-3}$\tabularnewline
\hline 
\end{tabular}%
\begin{tabular}{|c|c|c|}
\hline 
20\% sys. & \multicolumn{2}{c|}{$SS_{\mathrm{disc}}\geq5$}\tabularnewline
\hline 
\hline 
$\eta_{u}=\eta_{c}$ & $0.065$ & $1.10\times10^{-3}$\tabularnewline
\hline 
Only $\eta_{u}$ & $0.073$ & $1.39\times10^{-3}$\tabularnewline
\hline 
Only $\eta_{c}$ & $0.16$ & $6.63\times10^{-3}$\tabularnewline
\hline 
\end{tabular}
\par\end{raggedright}
\raggedright{}%
\begin{tabular}{|c|c|c|}
\hline 
Scenario & \multicolumn{2}{c|}{$SS_{\mathrm{exc}}\geq1.645$}\tabularnewline
\hline 
\hline 
$\eta_{u}=\eta_{c}$ & $0.034$ & $3.00\times10^{-4}$\tabularnewline
\hline 
Only $\eta_{u}$ & $0.036$ & $3.38\times10^{-4}$\tabularnewline
\hline 
Only $\eta_{c}$ & $0.12$ & $3.74\times10^{-3}$\tabularnewline
\hline 
\end{tabular}%
\begin{tabular}{|c|c|c|}
\hline 
20\% sys. & \multicolumn{2}{c|}{$SS_{\mathrm{exc}}\geq1.645$}\tabularnewline
\hline 
\hline 
$\eta_{u}=\eta_{c}$ & $0.036$ & $3.38\times10^{-4}$\tabularnewline
\hline 
Only $\eta_{u}$ & $0.040$ & $4.17\times10^{-4}$\tabularnewline
\hline 
Only $\eta_{c}$ & $0.12$ & $3.74\times10^{-3}$\tabularnewline
\hline 
\end{tabular}
\end{table}

\begin{acknowledgments}
No funding was received for this research.

The authors are grateful to Ulku Ulusoy for a careful reading of the
manuscript. We wish to acknowledge the support of the AUHEP group,
offering suggestions and encouragement. The numerical calculations
reported in this paper were partially performed at TUBITAK ULAKBIM,
High Performance and Grid Computing Center (TRUBA resources).
\end{acknowledgments}

\end{document}